\documentclass[11pt]{article}
\usepackage{tarz} 
\usepackage[dvips]{epsfig}
\usepackage{rotating}
\usepackage{caption}
\usepackage[utf8]{inputenc}
\usepackage{lscape}
\usepackage{natbib}
\usepackage[english]{babel}
\usepackage{amsmath}
\usepackage{amssymb}
\usepackage{soul}

\title{\textsc{\LARGE{COVID-19: What If Immunity Wanes?} 
}
	\vspace{0.5cm}
}	

\author{
	\textsc{\color{ChadBlue}{\Large{{M. Alper \c{C}enesiz}}}}
	\thanks{Nottingham Trent University. E-mail address: {alper.cenesiz@ntu.ac.uk}.}\and 
	\textsc{\color{ChadBlue}{\Large{{Lu\'{\i}s Guimar\~{a}es}}}}
	\thanks{Queen's University Belfast. E-mail address: {l.guimaraes@qub.ac.uk}.}
	\vspace{0.5cm}
}

\date{\small 	August 3, 2020} 

\usepackage{fancyhdr}
\begin{document}
	
	\maketitle
	\thispagestyle{fancy}
	
	\begin{spacing}{1}
	\begin{abstract}
		\textit{Using a simple economic model in which social-distancing reduces contagion, we study the implications of waning immunity for the epidemiological dynamics and social activity. If immunity wanes, we find that COVID-19 likely becomes endemic and that social-distancing is here to stay until the discovery of a vaccine or cure. But waning immunity does not necessarily change optimal actions on the onset of the pandemic. Decentralized equilibria are virtually independent of waning immunity until close to peak infections. For centralized equilibria, the relevance of waning immunity decreases in the probability of finding a vaccine or cure, the costs of infection (e.g., infection-fatality rate), and the presence of other NPIs that lower contagion (e.g., quarantining and mask use). In simulations calibrated to July 2020, our model suggests that waning immunity is virtually unimportant for centralized equilibria until at least 2021. This provides vital time for individuals and policymakers to learn about immunity against SARS-CoV-2 before it becomes critical. }
		
	\end{abstract}
	\end{spacing}
	\vspace{3mm}
	
	\newpage

\begin{spacing}{1}
\section{Introduction}
We do not know yet the duration of immunity against severe acute respiratory syndrome coronavirus 2 (SARS-CoV-2) causing coronavirus infectious disease 2019 (COVID-19). But early evidence points to waning immunity against SARS-CoV-2 \citep{Seow_etal_covid2020} and we know that immunity against other coronaviruses wanes within two years  \citep{Edridge_covid2020,Huang_covid2020,Kellam_Barclay_covid2020}. \\ 

If immunity against COVID-19 indeed wanes, then COVID-19 likely becomes endemic and herd immunity cannot be naturally reached. Therefore, ignoring waning immunity may lead to costly policies with irreversible consequences. Despite these risks, almost all the economics literature on the COVID-19 pandemic assumes permanent immunity.\footnote{In an already large and fast-growing economics literature addressing the COVID-19 pandemic, we are only aware of three papers allowing for waning immunity. We contrast our paper with these three papers below. The assumption of permanent immunity is also common outside of the economics literature:  e.g., \cite{Ferguson_covid2020report} and \cite{Wang_covid2020}.} Our paper fills this gap in the literature by assessing the implications of waning immunity for decentralized and centralized equilibria  in an economic model of an epidemic.\\

In the model, decision makers are constrained by disease contagion and maximize the difference between the utility from social activity and the cost of infection. The utility from social activity captures, in a stylized way, all the payoffs from economic and social actions that require physical proximity. Our approach is grounded in three reasons.\footnote{
	Among various approaches to study epidemics in economic models, ours follows \cite{Farboodietal_covid_2020}, \cite{Garibaldietal_covid2020}, \cite{Guimaraes_covid2020},  and \cite{Toxvaerd_covid2020} by directly modeling the choice of social activity. Another approach is to assume contacts are a function of the level and type of i) consumption \citep{Eichenbaum_Rebelo_Trabandt_covid2020_macroeconomics_of_epidemics, Eichenbaum_Rebelo_Trabandt_covid2020, Krueger_Uhlig_Xie_covid2020} and/or ii) labor \citep{Eichenbaum_Rebelo_Trabandt_covid2020_macroeconomics_of_epidemics, Eichenbaum_Rebelo_Trabandt_covid2020, Gloveretal_covid_2020}. Yet another approach is treating pandemics as exogenous shifts in state variables (e.g., human capital) \citep{Boucekkine_Diene_Azomahou_2008}. Such an approach resembles the MIT shock assessed by \cite{Guerrieri_Lorenzoni_Straub_Werning_covid2020} in the context of the COVID-19 pandemic. See also \cite{Philipson_Posner_aids_book_1993}, \cite{Kremer_aids_QJE_1996},  \cite{Chakraborty_Papageorgiou_Sebastian_JME_2010} and \cite{Greenwood_Kircher_Santos_Tertilt_aids_Etrica_2019} for an economic perspective of  HIV and malaria.} 
First, the main economic impact of the pandemic has been on sectors that rely on  physical proximity \citep{NBERw27431}. Second, there are also other significant costs of constrained social activity such as anxiety, distress, fatigue, and domestic violence \citep{NBERw27562,10.1093/qjmed/hcaa201}. Third, contagion of virus causing respiratory diseases is mostly unrelated with consumption and work \citep{ferguson2006, Eichenbaum_Rebelo_Trabandt_covid2020_macroeconomics_of_epidemics} but can be influenced by behavior. \\

The epidemiological dynamics in the model is  based on recurrence relations between three  (main) health states: susceptible (S), infected (I), and recovered (R) with the flow pattern $S\rightarrow I\rightarrow R\rightarrow S$ (and hence the conventional labeling SIRS).\footnote{The flow from recovered to susceptible stems from waning immunity.} An SIRS model nests both SIR and SIS models.\footnote{For an accessible review of epidemiological models, see \cite{Hethcote_2000}.} The canonical SIR model \citep{Kermack1927contribution} assumes that agents are permanently immune after they recover from the infection and is widely used in the economics literature addressing the COVID-19 pandemic \citep[e.g.,][]{Alvarezetal_covid_2020,Atkeson_covid2020,Eichenbaum_Rebelo_Trabandt_covid2020_macroeconomics_of_epidemics,Farboodietal_covid_2020}. The canonical SIS model assumes that agents are never immune and, thus, is employed in studying the economics of recurrent diseases \citep[e.g.,][]{Goenka_Liu_2012,Goenka_Liu_2019,Goenka_Liu_Nguyen_2014}.  An SIRS model is between an SIR and an SIS model by allowing agents to be immune but only temporarily. In light of the evidence on immunity against SARS-CoV-2 and other coronaviruses, an SIRS model is warranted to study the COVID-19 pandemic \citep{Kellam_Barclay_covid2020}.\\

In the canonical SIRS model, immunity is a binary variable:  agents are {either }immune or not. And after agents lose immunity they become as susceptible as any other susceptible agent. Waning immunity, however, does not necessarily mean that agents who lose immunity are as unprotected as those who were never infected \citep{Kuby_immunology, Huang_covid2020}.\footnote{In particular, \cite{Huang_covid2020} report that individuals can be infected with the same human coronaviruses one year after first infection but with lower severity.} Immunological memory (e.g., antibody count) might not be enough to avoid a reinfection but is likely enough for the body to react faster to a reinfection. For this reason,  our SIRS block allows susceptible agents to differ among themselves based on infection history. The heterogeneity in infection history can be captured by distinct i) probabilities of being infected, ii) recovery speed, iii) viral shedding, and iv) cost of infection. These possible distinctions are important as they may prevent an endemic COVID-19.\\

In Section \ref{sec_results}, we analyze the simplest case in which all susceptible agents, irrespective of their infection history, are alike. We reach two main conclusions. First, if immunological memory wanes, there is no vaccine or cure, and there is no major exogenous change in the contagiousness of the virus, then COVID-19 becomes endemic because of the  continuous flow of agents into the susceptible health state. In this scenario, both a social planner and decentralized individuals choose to social-distance forever. Second, the duration of immunity may not meaningfully change optimal choices in the initial months of the pandemic. We find that the decentralized equilibria is virtually independent of waning immunity for more than six months and until  close to peak infections because agents abstract from how their actions affect the probability that they are reinfected later. In slight contrast, we find that the centralized equilibria may vary with waning immunity depending on the costs of infection and the probability of finding a vaccine or cure. \\

An endemic COVID-19, induced by waning immunity, implies a higher present value of infection costs than a non-endemic one. In response to these higher costs, the social planner mandates further social-distancing. Yet, this extra social-distancing stemming from waning immunity  can be small  in the short run. If a vaccine is expected in 18 months and the costs of infection reflect an infection-fatality rate of 0.64\%, we find that optimal centralized policies are almost independent of waning immunity for more than 12 months. In this case, the short-term costs of infection are so high that the social planner severely constrains social activity to postpone those costs and wait for a vaccine. As social activity is already highly constrained, the marginal cost for society to further increase social-distancing is  huge. Thus, the social planner finds that the expected costs due to the endemic steady-state are of little relative importance in the short-term and almost does not react to them. In other words, when the short-term costs of the pandemic are very large, waning immunity is relatively unimportant at the early months of the pandemic. \\

If, on the other hand, the costs of infection are low (e.g., reflecting an infection-fatality rate close to 0.2\%), the costs of the pandemic are lower and the social planner mandates less social-distancing. As the costs are lower, the marginal cost of social-distancing are not prohibitively high,  giving the social planner room for maneuver  to act early to the prospect of the endemic steady-state.  Therefore, when immunity wanes, the social planner prefers to mandate relatively more social-distancing in the early months of the pandemic to reduce the costs of the endemic steady-state and gain time for a vaccine to arrive. Finally, a lower probability of discovering a vaccine increases the weight of future utility in the objective function in the same way as a lower discount factor does. This has an entirely different effect depending on waning immunity. When immunity is permanent, future utility is relatively high as the pandemic asymptotically disappears, which demotivates the social planner to postpone infections and mandate social-distancing. But, if immunity wanes in 10 months or two years on average, the present value of the costs of an endemic COVID-19 increase when a vaccine is expected to arrive late. Therefore, the social planner prefers to social-distance even more in the early days of the pandemic and act early to the problem of waning immunity. In sum, waning immunity only meaningfully changes centralized policies when the probability of discovering a vaccine is low or the societal marginal costs of acting early to the endemic steady-state are not unbearably high.\\

In Section \ref{sec_res_heterogeneity}, we analyze the case in which immunity wanes but susceptible agents differ based on infection history. Consistent with our previous results, if a vaccine is expected in 18 months and the costs of infections reflect an infection-fatality rate of 0.64\%, knowing whether susceptible agents differ based on infection history is not critical in the initial months of the epidemic. Furthermore, if agents that lost immunity are less likely to be infected or shed less virus, then COVID-19 does not become endemic. In this scenario, lower costs of infection and lower probability of finding a vaccine lead to markedly different choices in the short run. Thus, it is important to know whether immunity wanes and whether susceptible agents notably differ based on infection history. Finally, we find that susceptible agents that were immune can be excessively active from a social viewpoint, especially if they suffer much less from a reinfection, because they abstract from the risk of infecting others. Thus, policymakers should be aware of this extra source of risk if immunity wanes.\\

In a last set of simulations, in Section \ref{sec_today}, we change the starting date of the simulations. Our previous results are based on initial conditions matching the start of the COVID-19 pandemic. But, our results might differ as we are about six months past that in July 2020 and as other {non-pharmaceutical} interventions (NPIs) are in place besides social-distancing (e.g., mandatory mask use and quarantining of identified infected individuals). We find that, even if COVID-19 becomes endemic, the other NPIs in place allow for much more social activity. Furthermore, learning how the infection history affects the protection of individuals against reinfections becomes less important as contagion falls substantially. In fact, even a low probability of finding a vaccine or low costs of infection do not lead to markedly different centralized responses for many months. We conclude that individuals and policymakers have at least until 2021 to learn about the duration of immunity before it becomes truly important for decision making.\\

We are aware of three papers in the economics literature allowing for waning immunity: \cite{Eichenbaum_Rebelo_Trabandt_covid2020}, \cite{Giannitsarou_Kissler_Toxvaerd_covid2020}, and \cite{Malkov_covid2020}. These papers, however, differ from ours in crucial aspects including the object of study, approach, and modeling choices. \citeauthor{Eichenbaum_Rebelo_Trabandt_covid2020} study the role of testing and quarantines in a model with health state uncertainty and check the robustness of their findings if immunity wanes; thus, they do not fully explore how the duration of immunity affects contagion in the context of the current pandemic. \citeauthor{Malkov_covid2020}  focus on how waning immunity affects the epidemiological dynamics during the COVID-19 pandemic, but he does not allow individuals and the social planner to endogenously react in his simulations. \citeauthor{Giannitsarou_Kissler_Toxvaerd_covid2020} assess the centralized problem during the pandemic in case immunity wanes, but they do not contrast the centralized and decentralized equilibria and their results differ from ours due to modeling and calibration choices.\footnote{There are two other relevant differences. As \citeauthor{Eichenbaum_Rebelo_Trabandt_covid2020} and \citeauthor{Malkov_covid2020}, \citeauthor{Giannitsarou_Kissler_Toxvaerd_covid2020} assume that all susceptible agents are alike irrespective of infection history. And, as \citeauthor{Eichenbaum_Rebelo_Trabandt_covid2020}, \citeauthor{Giannitsarou_Kissler_Toxvaerd_covid2020} place their simulations at the start of the pandemic and assume that only one non-pharmaceutical intervention  is in place (testing in the case of \citeauthor{Eichenbaum_Rebelo_Trabandt_covid2020} and mandatory social-distancing in the case of \citeauthor{Giannitsarou_Kissler_Toxvaerd_covid2020}). Our last set of simulations in which we include the effects of other NPIs brings, thus, further insights to the current policy discussion.} In Section \ref{sec_discussion}, we contrast in more detail our results with those in the three papers.\\

\section{Model}\label{sec_model}
We build an economic model of an epidemic in which agents face a trade-off between social activity and exposure to the virus. This trade-off results from the link between the epidemiological  and utility-maximization blocks of the model. The link, in turn, stems  from  our assumption, following  \cite{Farboodietal_covid_2020},  \cite{Garibaldietal_covid2020}, and \cite{Guimaraes_covid2020}, that new infections depend on the number of susceptible and infected agents and the social activity chosen by susceptible agents. The model is set in discrete time. The population is constant and of measure one. We focus  on symmetric equilibria in which  agents with the same health state behave the same. With this and applying the law of large numbers, we do not separately denote individual and aggregate variables.\\

We distinguish  agents that become susceptible after recovery from agents that were never infected because the former, although no longer immune, may still have some immunological memory. The remaining immunological memory may allow for a lower probability of infection, faster recovery, lower viral shedding, and lower costs of infection. We refer to agents that were never infected as \textit{primary} and agents that were infected at least once as \textit{secondary}. To further ease our exposition, we use the index $j\in\{p,q\}$,  when referring primary and secondary agents, respectively.
\subsection{Epidemiological Block}
The population in period $t$ consists of five groups of agents: primary susceptible, $s_{p,t}$, primary infected, $i_{p,t}$, recovered, $r_t$, secondary susceptible, $s_{q,t}$, and secondary infected, $i_{q,t}$. The number of new infections for each type is given by 
\begin{gather*}
\beta_j a_{j,t} s_{j,t} i_t,
\end{gather*} where $\beta_j$ is the measure of contagiousness for susceptible agents of type $j$ with $\beta_q\leq\beta_p$, $a_{j,t}\in[0,1]$ is the social activity of susceptible agents of type $j$,  and 
\begin{gather}
i_t = i_{p,t} + \sigma i_{q,t}\label{define_I}
\end{gather}
is the number of infected agents. We adjust $i_{q,t}$ with $\sigma\leq1$ to allow secondary infected individuals shedding  less virus than primary infected ones.\\

The laws of motion governing the transitions between health states are the following:
\begin{alignat}{3}
& \label{eq2}  s_{p,t+1} = (1- \beta_p a_{p,t}i_t)s_{p,t},&\\
& i_{p,t+1} = \beta_p a_{p,t}s_{p,t}i_t + (1-\gamma_p) i_{p,t},&\\
& r_{t+1} = {\textstyle \sum\nolimits_j}\gamma_j i_{j,t}+(1- \alpha) r_t,&\\
& s_{q,t+1} = \alpha r_t+(1- \beta_q a_{q,t}i_t)s_{q,t},&\\
& i_{q,t+1} = \beta_q a_{q,t}s_{q,t}i_t + (1- \gamma_q) i_{q,t},&\label{law_N2i}
\end{alignat}
where $\gamma_j$ is the probability that an infected individual of type $j$ recovers and $\alpha$ is the probability that a recovered individual loses immunity. If $\alpha = 0$ and $a_{p,t}=1$ for all $t$, the model reduces to the canonical SIR model. If $\alpha>0$, $\sigma=1$, $\beta_p=\beta_q$, $\gamma_p=\gamma_q$, and $a_{j,t}=1$ for all $j$ and $t$,  the model reduces to the canonical SIRS model.\footnote{Under permanent immunity, $\alpha = 0$, the number of secondary susceptible agents remains zero. Under waning immunity, $\alpha >0$, with $\sigma=1$, $\beta_p=\beta_q$,  and $\gamma_p=\gamma_q$, identifying secondary agents is trivial.}
\subsection{Decentralized Problem}	
\subsubsection{Utility Maximization}\label{decentralized_problem}
In this section, we detail the lifetime utility maximization problem of a primary susceptible agent. Agents derive utility from their social activity. The utility function, denoted by $u(a)$ is single-peaked and its maximum is normalized to zero at $a=1$. The maximization problem of a primary susceptible agent  is given by
\begin{gather}
\max_{\{a_{p,t}, a_{p,t}\}_{t=0}^\infty}\sum\nolimits_{t=0}^{\infty}\sum\nolimits_j\Lambda^t\Big(s_{j,t}u(a_{j,t}) - \gamma_j\kappa_j i_{j,t}\Big),\nonumber
\end{gather}
subject to Eqs. (\ref{eq2}--\ref{law_N2i}). In this maximization problem, the fraction of  agents in each health state group corresponds to the (subjective) probability of the agent being in that state; $\Lambda$ is the discount factor; and $\kappa_j$ captures all the costs of recovering from the infection. As primary and secondary infected agents may respond differently to the infection (e.g., differ in symptoms severity), we set $\kappa_q\leq\kappa_p$. The decentralized optimum social activity is, then, governed by
\begin{alignat}{1}
&u'(a_{j,t}) = \beta_j i_t(V_{s_j,t}-V_{i_j,t}),\label{foc_dec_a1}\\
&\frac{V_{s_j,t}}{\Lambda} = u(a_{j,t+1}) + V_{s_j,t+1} - \beta_j a_{j,t+1}i_{t+1}(V_{s_j,t+1}-V_{i_j,t+1}),\\
&\frac{V_{i_j,t}}{\Lambda}  = V_{i_j,t+1} - \gamma_j(\kappa_j+V_{i_j,t+1} - V_{r,t+1}),\\
&\frac{V_{r,t}}{\Lambda}   = V_{r,t+1} + \alpha(V_{s_q,t+1}-V_{r,t+1}),\label{foc_dec_Nr}
\end{alignat}
for  both $j\in\{p,q\}$ and $V_{x,t}$ denotes the (shadow) value of the agent in state $x\in\{s_p,$ $s_q,$ $i_p,$ $i_q,$ $r\}$. Eq. \eqref{foc_dec_a1}  summarizes the trade-off.  Its left-hand side is the marginal utility of social activity while its right-hand side is expected marginal costs resulting from the possibility of infection. Marginal costs depend on how likely they are exposed by marginally increasing activity, $\beta_j i_t$. And it also depends on the change in the value caused by exposure, which is always positive,  $V_{s_j,t}-V_{i_j,t}>0$. Thus, susceptible agents restrain their social activity, $a_{j,t}\leq1$, to reduce exposure risk.\\

Eqs. (\ref{foc_dec_a1}-\ref{foc_dec_Nr}), determining the behavior of primary agents, are symmetric along $j$. Given that these equations do not depend on the (subjective) probability of being in any health state, the same equations also determine the behavior of secondary agents. Therefore, for brevity, we do not present the utility maximization problem of secondary agents. 
\subsubsection{Decentralized Equilibrium}
A decentralized equilibrium corresponds to a path of social activities, $\{a_{p,t}, a_{q,t}\}$, the number of infected agents, $i_t$, state variables, $\{s_{p,t}$, $s_{q,t}$, $i_{p,t}$, $i_{q,t}$, $r_t\}$, and shadow values, $\{V_{s_p,t}$, $V_{s_q,t}$, $V_{i_p,t}$, $V_{i_q,t}$, $V_{r,t}\}$, that satisfy Eqs. (\ref{define_I}--\ref{foc_dec_Nr}).
\subsection{Centralized Problem}
\subsubsection{Utility Maximization}
In this section, we present the maximization problem of the social planner. The social planner chooses socially optimal activity by directly influencing aggregate variables. In particular, the maximization problem of the social planner is given by 
\begin{gather}
\max_{\{a_{p,t}, a_{p,t}\}_{t=0}^\infty}\sum\nolimits_{t=0}^{\infty}\sum\nolimits_j\Lambda^t\Big(s_{j,t}u(a_{j,t}) - \gamma_j\kappa_j i_{j,t}\Big),\nonumber
\end{gather}
subject to Eqs. (\ref{define_I}-\ref{law_N2i}). Relative to the decentralized problem, Eq. \eqref{define_I} is the additional constraint because the social planner internalizes how infected individuals affect contagion. The socially optimum social activity is, then, governed by
\begin{alignat}{2}
&u'(a_{j,t}) = \beta_j i_t(V_{s_j,t}-V_{i_j,t}),\label{foc_cen_a1}\\
&\frac{V_{s_j,t}}{\Lambda} = u(a_{j,t+1}) + V_{s_j,t+1} - \beta_j a_{j,t+1}i_{t+1} (V_{s_j,t+1}-V_{i_j,t+1}),\\
&\frac{V_{i_j,t}}{\Lambda} = V_{i_j,t+1} - \gamma_j(\kappa_j+V_{i_j,t+1} - V_{r,t+1})- \sigma_j{\textstyle \sum\nolimits_j}\beta_j a_{j,t+1}s_{j,t+1}(V_{s_j,t+1}-V_{i_j,t+1})\\
&\frac{V_{r,t}}{\Lambda} =  V_{r,t+1} + \alpha(V_{s_q,t+1}-V_{r,t+1}),\label{foc_cen_Nr}
\end{alignat}
for  both $j\in\{p,q\}$, 
and $\sigma_j=\begin{cases}

1 & \text{if } j=p,\\

\sigma & \text{if } j=q.

\end{cases}$ \quad Comparing this set of equations governing the optimal choice of the social planner with that governing the optimal choice of agents in the decentralized problem (Eqs. \ref{foc_dec_a1}-\ref{foc_dec_Nr}), we can see that the only difference is in the shadow values of the infected states. This difference reflects a key externality emphasized in the literature: in a decentralized equilibrium, agents decide their social activity without considering the risk of infecting others. As a result, both $V_{i_p,t}$ and $V_{i_q,t}$ are lower in the social planner's problem, which (\textit{ceteris paribus}) restrains social activity relative to the decentralized equilibrium. Part of our objective in this paper is to analyze how the possibility of recovered agents losing immunity distances decentralized and centralized choices.
\subsubsection{Centralized Equilibrium}
A centralized equilibrium corresponds to a path of social activities, $\{a_{p,t}, a_{q,t}\}$, the number of infected agents, $i_t$, state variables, $\{s_{p,t}, s_{q,t}, i_{p,t}, i_{q,t}, r_t\}$, and shadow values, $\{V_{s_p,t}$, $V_{s_q,t}$, $V_{i_p,t}$, $V_{i_q,t}$, $V_{r,t}\}$, that satisfy Eqs. (\ref{define_I}-\ref{law_N2i}) and Eqs. (\ref{foc_cen_a1}-\ref{foc_cen_Nr}).
\section{Calibration}\label{sec_calibration}
We summarize our parameter choices in Table \ref{benchmark parameters}. Each period in the model corresponds to one day. The discount factor includes both a time discount rate, $\rho$, and the probability of finding a cure-for-all, $\delta$, that would end the problem. In particular, we set $\Lambda=\frac{1}{1+\rho}\frac{1}{1+\delta}$, $\rho=0.05/365$, and $\delta=0.67/365$ reflecting a yearly discount rate of $5\%$ and the probability of finding the cure-for-all of $67\%$ within a year \citep[see, e.g.,][]{Alvarezetal_covid_2020, Farboodietal_covid_2020}.
\begin{center}
	\begin{table}[h]
		\caption{Benchmark  Calibration} 
		\begin{center}
			\begin{footnotesize}
				\begin{tabular}[h]{llr}
					\hline
					\hline
					{}&{}&{}\\
					{Discount factor:}&{}&$\Lambda=\frac{1}{1+0.05/365}\frac{1}{1+0.67/365}$ \\
					{Cost of infection:}&{}&$\kappa_p=\kappa_q=512$ \\
					{Average number of days as infected:}&{}&$\gamma_p^{-1}=\gamma_q^{-1}=18$ \\
					{Infectiousness:}&{}&$\beta_p=\beta_q=2.4/18$ \\
					{Average number of days immune:}&{}&$\alpha^{-1}=750$ \\
					{Relative viral shedding of secondary infected:}&{}&$\sigma=100\%$ \\
					{}&{}&{}\\
					\hline\hline
					\multicolumn{3}{c}{}\\\label{benchmark parameters}
				\end{tabular}
			\end{footnotesize}
		\end{center}
	\end{table}
\end{center}

As in \cite{Farboodietal_covid_2020} and \cite{Guimaraes_covid2020}, the utility of social activity is determined by:
\begin{gather}
u(a) = \log(a) - a +1,\label{utils}
\end{gather}
which guarantees that $u(a)$ is single-peaked with maximum at $a=1$ and $u(1)=u'(1)=0$. We also closely follow \citeauthor{Farboodietal_covid_2020} to find the cost of infection, $\kappa_p$. Assuming that the value of life is US$\$10$ million and assessing how much agents would be willing to permanently reduce their consumption to permanently lower the probability of dying by $0.1\%$, we find that the value of life is $80000$ in model units. Based on this, we obtain $\kappa_p$ using the probability of dying conditional on infection. The meta-analysis of \cite{Meyerowitz-Katz_Merone_covid2020} suggests that about $0.64\%$ of those infected with the virus, die. Thus, we set $\kappa_p=512$. \\

We follow \cite{Atkeson_covid2020} and most of the economics literature assessing the COVID-19 pandemic and assume that infected individuals remain so  for 18 days, $\gamma_p^{-1}=18$. To calibrate $\beta_p$, we follow \cite{Acemoglu_Chernozhukov_Werning_Whinston_covid2020} and assume $\beta_p=2.4/18$, implying a basic reproduction number, $R0$, of $2.4$. This number is relatively optimistic in light of, for example, the $R0$ assumed in \cite{Alvarezetal_covid_2020} of $3.6$. \\

At this stage, the duration of immunity against COVID-19 and how secondary agents differ from primary agents is unknown. To calibrate the probability that recovered individuals lose immunity, we use the evidence regarding other coronaviruses surveyed in \cite{Huang_covid2020} and also the assumption in \cite{Eichenbaum_Rebelo_Trabandt_covid2020} and set $\alpha^{-1}=1/750$, implying that agents have immunity for about two years. Regarding the remaining parameters, in our benchmark we simply assume that $\beta_q=\beta_p$, $\gamma_q=\gamma_p$, $\kappa_q=\kappa_p$ and $\sigma=100\%$. Therefore, our benchmark calibration implies a SIRS model augmented with the endogenous choice of social activity.\\

We solve the model using a shooting algorithm as detailed in \cite{Garibaldietal_covid2020}. As a starting point, we assume that 1 in a million agents are primary infected, $i_{p} = 1/10^6$, and the remaining are primary susceptible.
\section{Results}\label{sec_results}
\subsection{Main Results}\label{sec_main_results}
\begin{figure}[!htbp]
	\centering
	\caption{\linespread{1.0}\selectfont 
		\label{fig_ups1}%
		The Role of Immunity Duration
		\vspace*{0.1cm}
	}
	\includegraphics[scale=0.7]{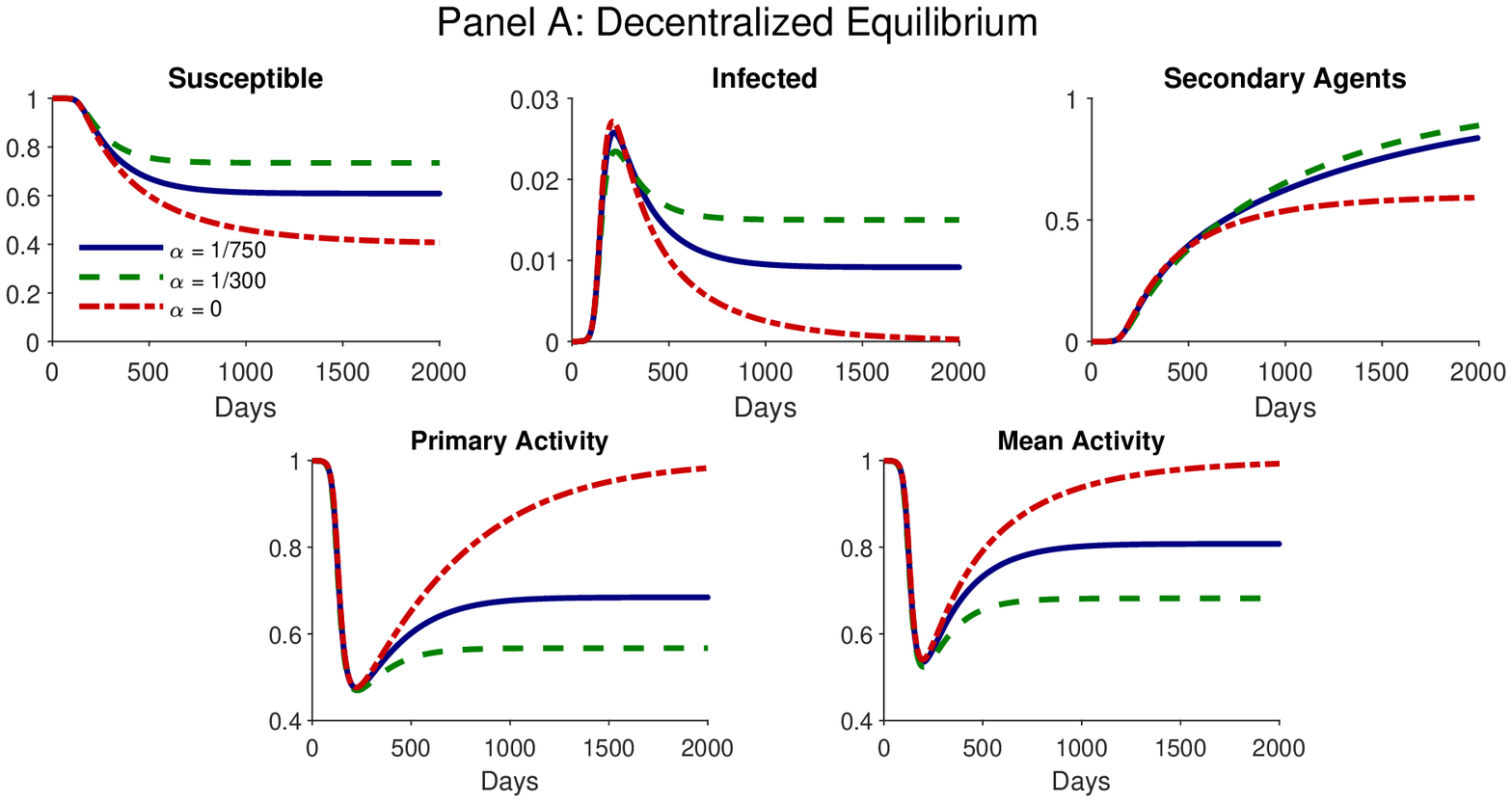}\vspace{0.35cm}
	\includegraphics[scale=0.7]{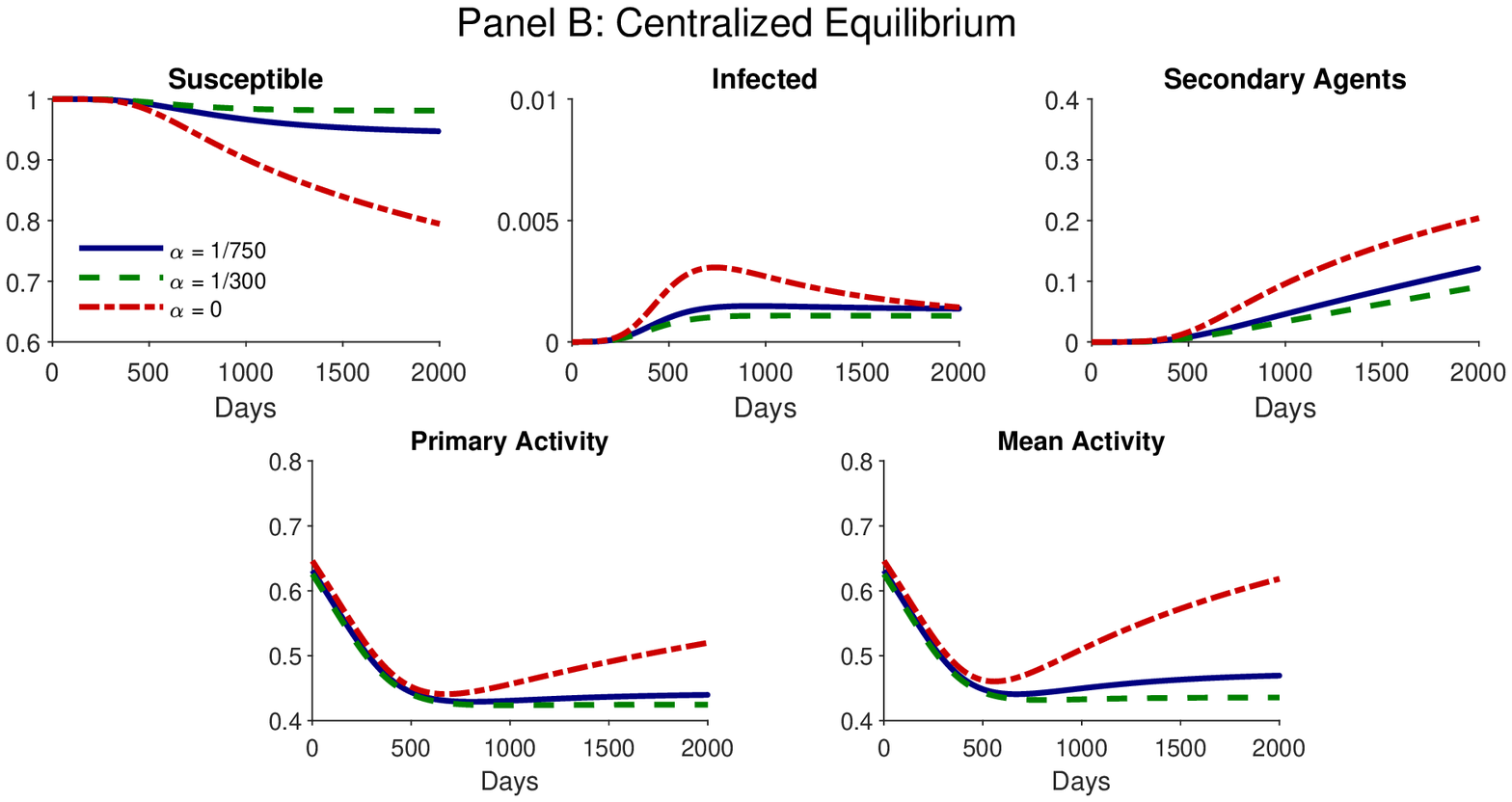}
	\parbox{13cm}%
	{\vspace*{.5cm}
		\linespread{1.0}\selectfont 
				\footnotesize \textit{Note:} Susceptible agents are $s_{p,t} + s_{q,t}$; infected agents are $i_{p,t} + i_{q,t}$; secondary agents are $s_{q,t} + i_{q,t} + r_t$; primary activity is $a_{p,t}$ (which, in this case, equals secondary activity, $a_{q,t}$); and mean activity is $s_{p,t}a_{p,t} + s_{q,t}a_{q,t} + i_{p,t} + i_{q,t} + r_t$.
	}%
\end{figure}

Panels A and B of Figure \ref{fig_ups1} present how a waning immunological memory affects optimal decentralized and centralized dynamics, respectively. The blue (solid) lines assume our benchmark, i.e., agents are immune for two years on average. The green (dashed) lines assume, as a lower bound and consistent with \cite{Huang_covid2020} and \cite{Kissler_etal_covid2020}, that immunological memory lasts only 10 months. The red (dot-dashed) lines assume, as an upper bound and as in the economics literature (e.g. \citealp{Alvarezetal_covid_2020, Eichenbaum_Rebelo_Trabandt_covid2020_macroeconomics_of_epidemics, Farboodietal_covid_2020}), that immunity lasts forever (implying an SIR model).\\

Two of our findings in Figure \ref{fig_ups1} are striking. First, if immunological memory wanes ($\alpha>0$), then  in both centralized   and decentralized equilibria, social activity is severely and permanently curtailed until the discovery of a vaccine or cure. This results from the continuous flow of agents from immune to susceptible, implying a continuous flow from susceptible to infected and, therefore, a permanent exposure risk. Thus, if immunity wanes, COVID-19  reaches an endemic steady-state. In the centralized equilibrium, social activity stabilizes at about 55\% lower than absent the epidemic. In the decentralized equilibrium, social activity reaches its minimum  after about 200 days and then recovers slightly to its long run value, 30\% lower than absent the epidemic. If agents never lose immunity, $\alpha=0$, the results are very different.  In this case, all agents return to normal activity as infections asymptotically disappear. This happens faster, albeit at a higher social cost, in the case of the decentralized equilibrium, leading to much higher peak infections. Furthermore, in the decentralized equilibrium, approximately 60\% of the agents are infected at least once within three years, which differs substantially from about 5\% in the centralized equilibrium.\\

Second, the underlying  duration of immunity barely moves the initial dynamics of epidemiological variables and social activity for around 200 days in the decentralized and 400 days in the centralized equilibrium. This result is partly explained by the low accumulation of secondary agents as few agents obtain and lose immunity in the initial months of the epidemic even when immunity wanes after 10 months, $\alpha=1/300$. But other factors play important roles, especially in centralized equilibria. In decentralized equilibria, agents do not take into account how their actions, by affecting infections, change the pace at which they might be reinfected. As a result, social activity in decentralized equilibria is mostly affected by the dynamics of infected agents. As soon as many agents start losing immunity and become susceptible and infected again, the effects of waning immunity become visible in optimal social activities.\\

In centralized equilibria, however, the externalities of social activity are considered in decision-making. The social planner knows that by reducing social activity, it lowers and postpones infections and, thereby, decreases the number of secondary agents that lose immunity. Furthermore, the social planner is aware of the costs of the endemic steady-state. These two factors combined motivate the social planner to constrain social activity by more when waning immunity induces an endemic COVID-19. Yet, surprisingly, in our benchmark case, the optimal centralized social activity is almost unmoved by the duration of immunity for 400 days. \\

The social planner aims to minimize the sum of the present value of the costs of infection and of social-distancing. If immunity is permanent, Panel B of Figure \ref{fig_ups1} shows that the best option to minimize social costs is to endure high social-distancing, postpone infections, and wait for the vaccine. If, on the other hand, immunity wanes, future infection costs increase but their present value is substantially discounted because the vaccine or cure is expected in 18 months. Furthermore, as social activity is heavily constrained even if immunity is permanent, the marginal costs of social-distancing are high and very sensitive to further increases in social-distancing due to the curvature of the utility function. Put differently, the social planner lacks room to maneuver to strongly react to waning immunity in the early months of the pandemic. These two factors combined explain why waning immunity is relatively unimportant for many months in determining optimal social-distancing.\\

\begin{figure}[!htbp]
	\centering
	\caption{\linespread{1.0}\selectfont 
		\label{fig_ups2}%
		The Role of Immunity Duration - Centralized Equilibria
		\vspace*{0.1cm}
	}
	\includegraphics[scale=0.7]{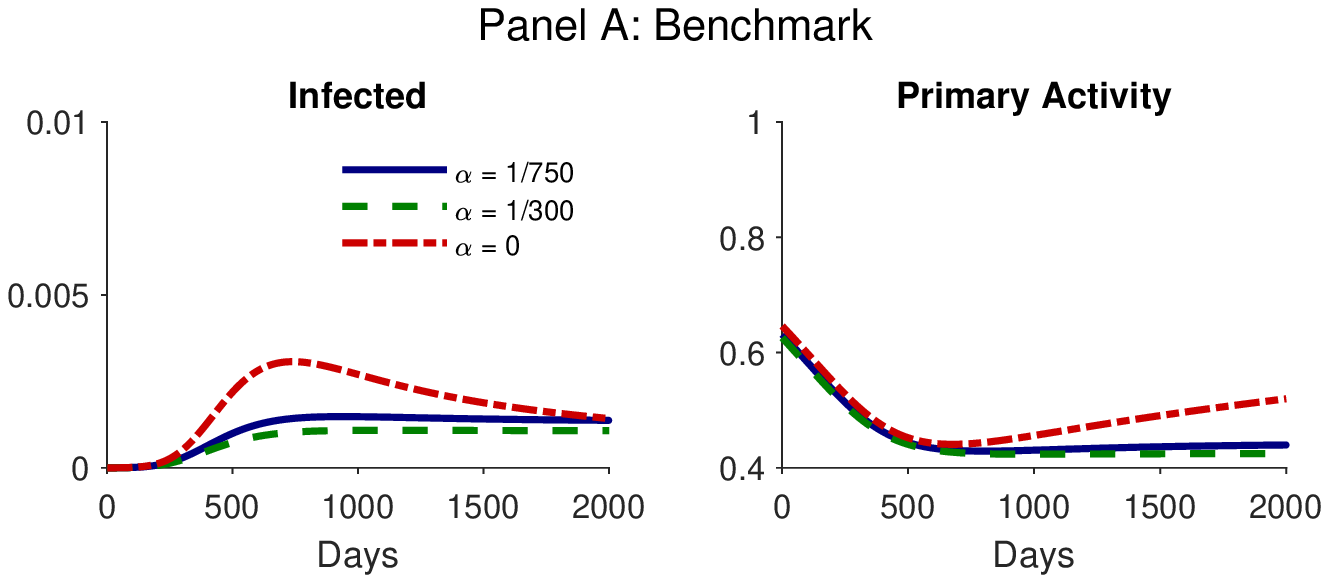}\vspace{0.35cm}
	\includegraphics[scale=0.7]{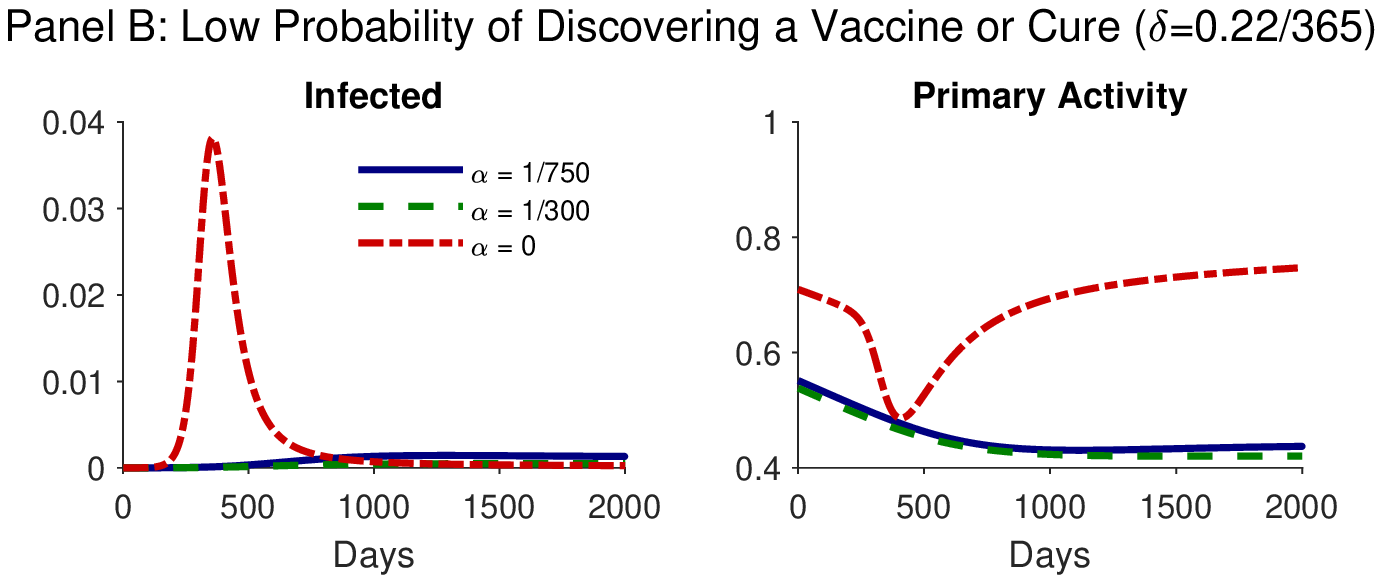}\vspace{0.35cm}
	\includegraphics[scale=0.7]{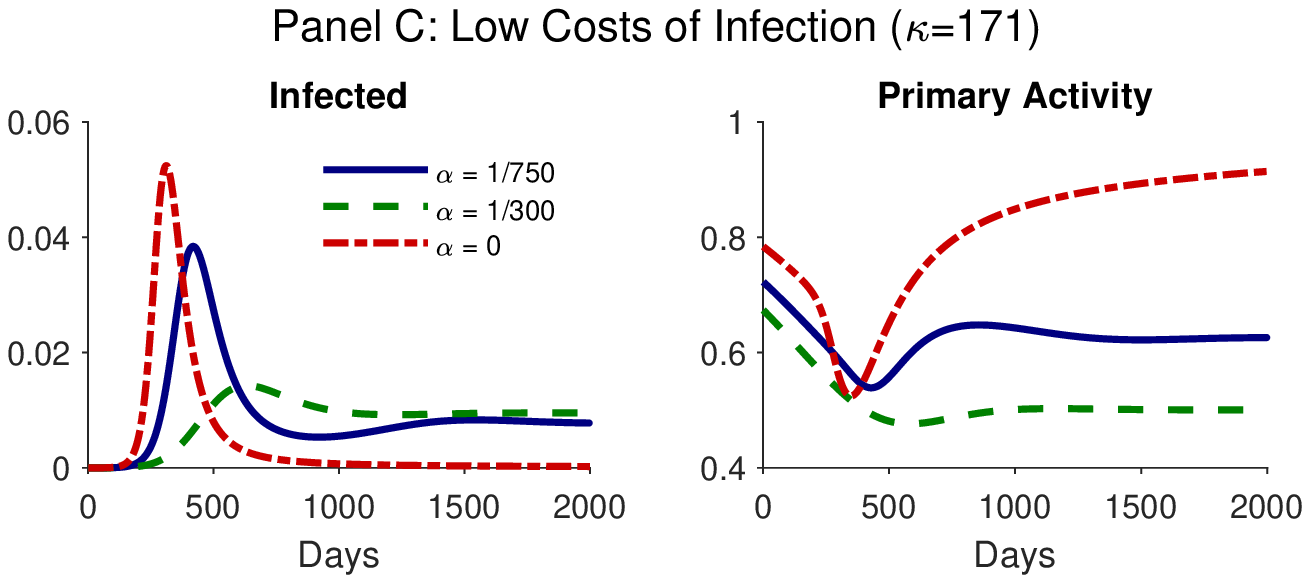}
	\parbox{13cm}%
	{\vspace*{.5cm}
		\linespread{1.0}\selectfont 
		\footnotesize \textit{Note:} Infected agents are $i_{p,t} + i_{q,t}$; primary activity is $a_{p,t}$ (which, in this case, equals secondary activity, $a_{q,t}$).
	}%
\end{figure}

To gain further insight, in Figure \ref{fig_ups2}, we show how two key parameters change the number of infected agents and social activity of primary agents in centralized equilibria. 
Panel A depicts again the benchmark cases to ease comparison. Panel B depicts the results when expected time to find  a vaccine or cure is 4.5 years, implying $\delta$  is a third of its benchmark value.  Panel C depicts the results when the infection-fatality rate is approximately 0.21\%, implying $\kappa_j$ is a third of its benchmark value. This figure shows that waning immunity matters in these two deviations from benchmark in the centralized equilibria.\footnote{Figure \ref{fig_7} in the Appendix shows that the way waning immunity affects decentralized equilibria relies much less on $\kappa_p$ and $\delta$.} The results are particularly staggering in the case of low $\delta$: in this scenario, peak infections occur much earlier and is more than 20 times higher when immunity is persistent than when immunity wanes.\\

A lengthier period to discover a vaccine or cure, captured by a lower $\delta$, implies that the social planner must restrict social activity for more time to avoid infections and wait for the vaccine or cure. We find that the corresponding increase in the present value of social-distancing costs greatly exceeds the increase in the present value of the costs of infection if immunity is permanent. Therefore, the social planner allows for more infections. The opposite holds when immunity wanes. Technically, a lower $\delta$ reduces the discount factor, increasing the present value of the infection costs caused by waning immunity and the endemic COVID-19. Therefore, the social planner reacts even stronger to the pandemic when it emerges if immunity wanes and the vaccine is expected later in time. \\

A reduction in the infection-fatality rate, captured by a lower, $\kappa_j$, implies less costs of infection and, thus, more social activity whatever is $\alpha$. But the rise in social activity increases in the duration of immunity (i.e., decreases in $\alpha$). When the costs of infection, $\kappa_j$, are lower, the implied  point of the reduced social activity is under the flatter range of the curved utility.\footnote{In an experiment (not reported), we varied the curvature of the utility function and find  that the changes in social activity brought by waning immunity decrease in the curvature.} Thus, the marginal cost of additional social-distancing is also relatively low, increasing the room to maneuver of the social planner.  Therefore, the social planner acts stronger from the onset of the pandemic to reduce the costs of an endemic COVID-19 and gain time for the discovery of a cure or vaccine. This difference in optimal choices lead to clearly different disease dynamics: the faster immunity wanes, the more the social planner postpones and reduces peak infections.\\

In sum, waning immunity implies a persistent reduction in social activity either individually chosen or mandated. But because individuals lack altruism, implying a weaker link between choice and (re)infection, the early response to the pandemic in decentralized equilibria is not dependent on waning immunity. In centralized equilibria, however, waning immunity may affect the early response to the pandemic depending on the magnitude of the costs of infection and critically on how likely a vaccine or cure is expected to arrive. Yet, in our benchmark calibration, which we find plausible, waning immunity barely affects early optimal choices of social activity in the centralized equilibria.

\subsection{Discussion}\label{sec_discussion}
In this section, we contrast our findings with the three papers in the economics literature that study waning immunity. \cite{Eichenbaum_Rebelo_Trabandt_covid2020} study the role of testing and quarantining in a model linking  consumption and labor choices to contagion. They also find that decentralized individuals permanently reduce their activity (consumption and labor supply) due to the endemic steady-state caused by waning immunity. Furthermore, their Figure 9 suggests that, for over a year, waning immunity is virtually irrelevant for decentralized decisions.  Yet, waning immunity affects their centralized equilibria in a way different from ours because of the different policy instruments considered. Their testing and quarantining polices rule out endemic steady states because asymptotically all individuals are continuously tested and infected ones are quarantined. Therefore, waning immunity neither restricts social planner's actions nor permanently constrains economic activity in \cite{Eichenbaum_Rebelo_Trabandt_covid2020}.\\

\cite{Giannitsarou_Kissler_Toxvaerd_covid2020} study the effects of waning immunity on social-distancing policy. Notable differences between our paper and theirs are as follows. They assume that the pandemic ends in six years (by the discovery of a vaccine), ruling out any endemic steady state.  As a result, social activity returns to normal in their simulations. Moreover, the costs of infection and social-distancing are much lower in their model. They assume that the costs of infection are  10\% lower output by infected and zero output by deceased only for the time span of the pandemic. The costs of social-distancing are quadratic and finite in a mandated full-lockdown, which provide a vast room to maneuver for the social planner to act. Therefore, when immunity wanes, they obtain deferment of peak infections and a negative relation between immunity duration and mandated social distancing (similar to our results in the low cost of infection case, Figure \ref{fig_ups2}, Panel C). \\

\cite{Malkov_covid2020} studies how waning immunity affects the dynamics of an epidemiological model under different calibrations of the basic reproduction number. He concludes that until close to peak infections, waning immunity barely changes the disease dynamics. Although \citeauthor{Malkov_covid2020} does not include endogenous decision making in his model, his findings are relatively close to our findings in the decentralized equilibria as waning immunity also only matters close to peak infections. But his findings differ substantially from our results in the centralized equilibria. In this case, the social planner takes into account the future costs of waning immunity in his early response to the pandemic, which in turn, leads to different disease dynamics.

\section{Heterogeneous Susceptible Agents}\label{sec_res_heterogeneity}
\begin{figure}[!htbp]
	\centering
	\caption{\linespread{1.0}\selectfont 
				\label{fig_robust1}%
	Heterogeneous Susceptible Agents
		\vspace*{0.1cm}
	}
	\includegraphics[scale=0.7]{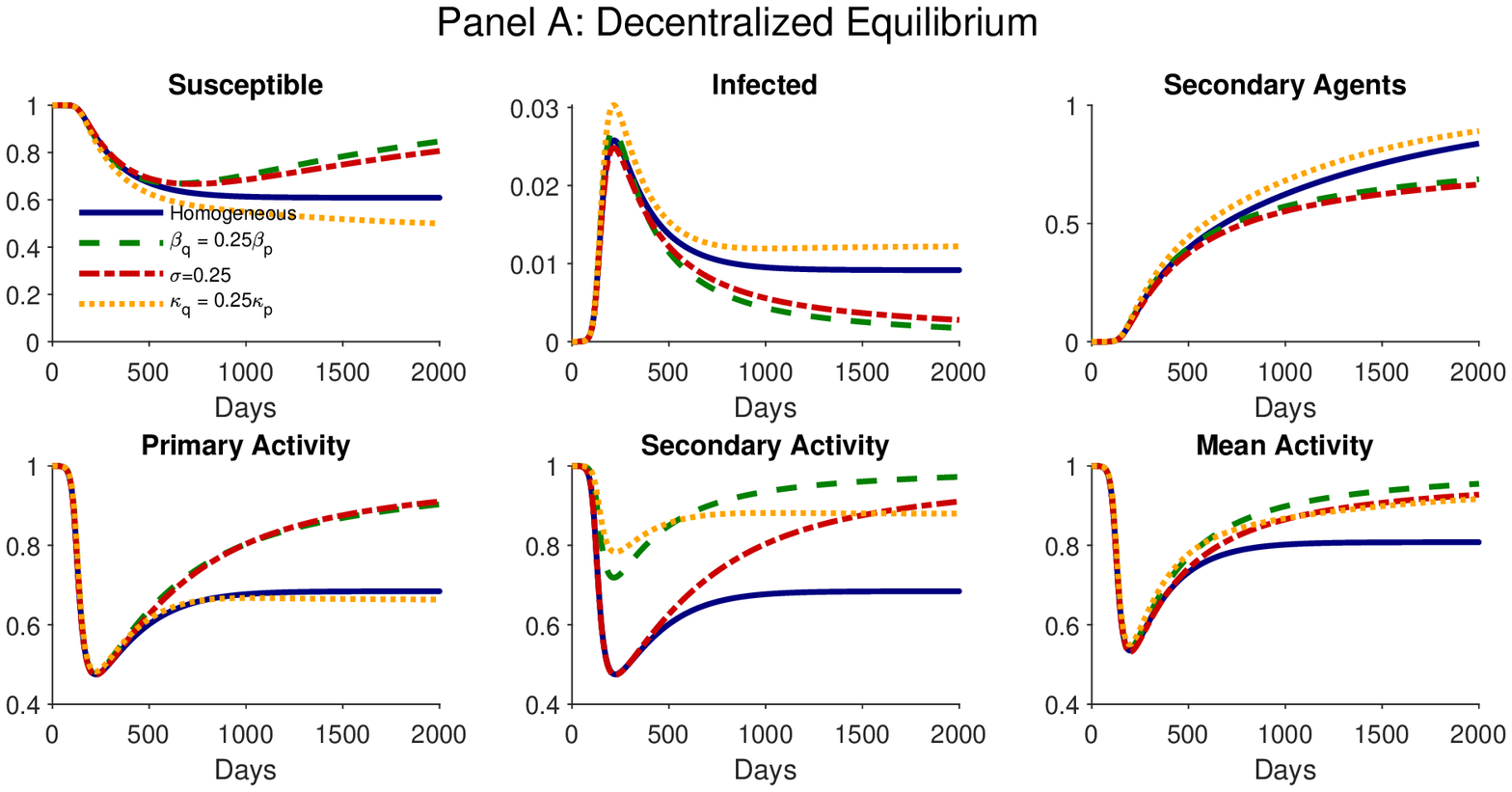}\vspace{0.35cm}
	\includegraphics[scale=0.7]{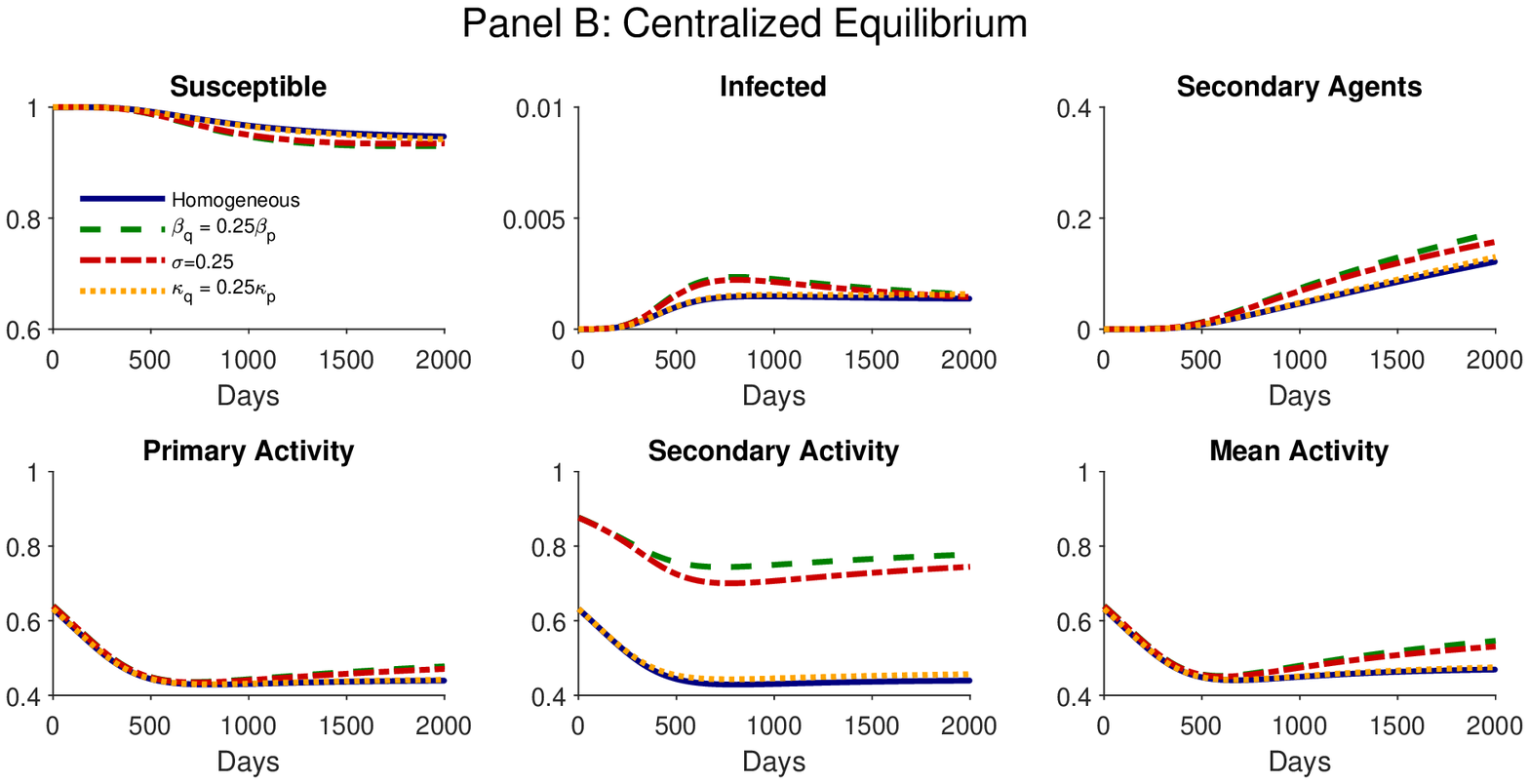}
	\parbox{13cm}%
	{\vspace*{.5cm}
		\linespread{1.0}\selectfont 
				\footnotesize \textit{Note:} Homogeneous refers to the case in which secondary and primary agents are alike. Susceptible agents are $s_{p,t} + s_{q,t}$; infected agents are $i_{p,t} + i_{q,t}$; secondary agents are $s_{q,t} + i_{q,t} + r_t$; primary activity is $a_{p,t}$; secondary activity is $a_{q,t}$; and mean activity is $s_{p,t}a_{p,t} + s_{q,t}a_{q,t} + i_{p,t} + i_{q,t} + r_t$.
	}%
\end{figure}

\begin{figure}[!htbp]
	\centering
	\caption{\linespread{1.0}\selectfont 
				\label{fig_robust2}%
		Heterogeneous Susceptible Agents - Centralized Equilibria
		\vspace*{0.1cm}
	}
	\includegraphics[scale=0.7]{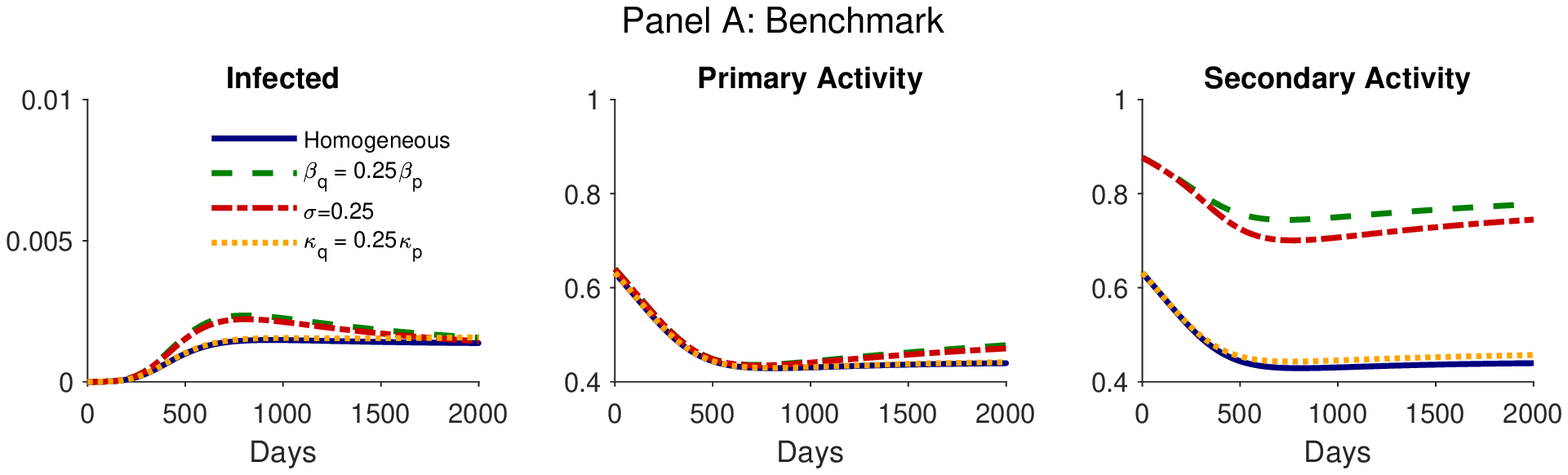}\vspace{0.35cm}
	\includegraphics[scale=0.7]{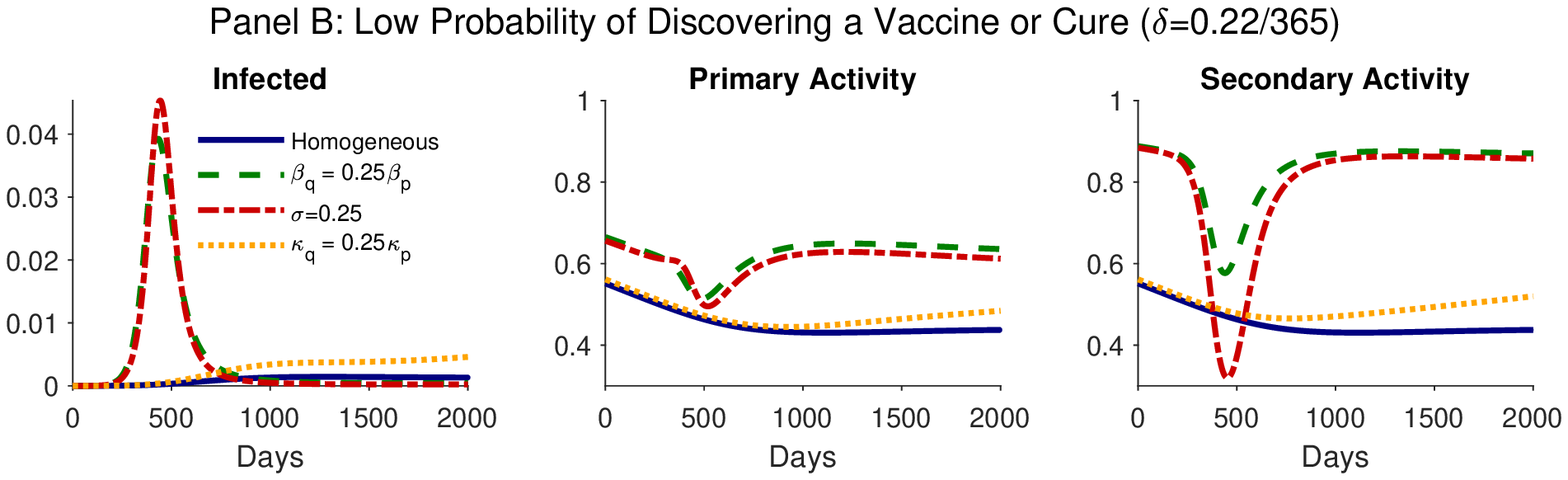}\vspace{0.35cm}
	\includegraphics[scale=0.7]{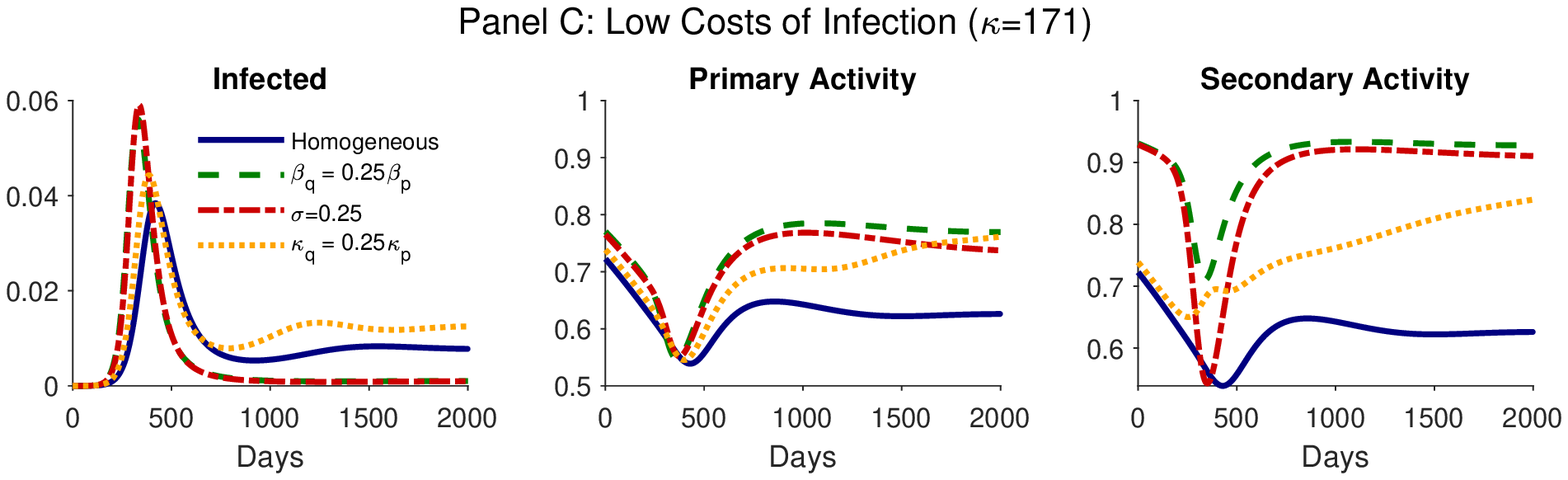}
	\parbox{13cm}%
	{\vspace*{.5cm}
		\linespread{1.0}\selectfont 
				\footnotesize \textit{Note:} Homogeneous refers to the case in which secondary and primary agents are alike. Infected agents are $i_{p,t} + i_{q,t}$; primary activity is $a_{p,t}$; secondary activity is $a_{q,t}$.
	}%
\end{figure}

So far,  we have analyzed an SIRS model augmented with endogenous social activity. Using our benchmark calibration, in Figure \ref{fig_robust1}, we illustrate how our results change when secondary susceptible and infected agents differ from their primary counterparts in three  aspects. Figure \ref{fig_robust2} complements our illustration in Figure \ref{fig_robust1} by showing how our results differ if $\delta$ and $\kappa_p$ are low. Green (dashed) lines show the case in which secondary susceptible agents are $75\%$ less likely to be infected than primary susceptible agents;\footnote{This implies a reduction of 75\% in $R0$. Different combinations of changes in $\beta_j$ and $\gamma_j$ leading to the same fall in $R0$ imply similar results.} red (dot-dashed) lines show the case in which secondary infected individuals shed 75\% less virus than primary infected;  yellow (dotted) lines show the case in which the costs of infection are 75\% lower for secondary agents; and blue (solid) lines show the benchmark. In the first two cases,  even though all agents eventually lose immunity,  asymptotic $R0$ is below 1 and, thus, the epidemic will asymptotically disappear as secondary agents gradually replace primary agents. In the case of $\kappa_q=0.25\kappa_p$, the cost of a reinfection is much lower but the flows between states do not asymptotically converge to zero.  That is, asymptotically, individuals are continuously infected but suffering much less than in the beginning of the epidemic. In this case, COVID-19 converges to an endemic steady-state, which is similar to that of other coronaviruses giving rise to flu-like symptoms \citep{Edridge_covid2020,Huang_covid2020,Kellam_Barclay_covid2020}.\\

Figure \ref{fig_robust1} and Panel A in Figure \ref{fig_robust2} show that if secondary and primary agents differ, there are little changes to the optimal social activity of primary susceptible agents for approximately a year and a half in both centralized and decentralized equilibria. This contributes to a similar path for the number of susceptible (both primary and secondary) agents for many months. Thus, as in the previous section, our benchmark calibration implies that any uncertainty caused by waning immunity is not much relevant for several months after the start of the epidemic.\\

Our results in the centralized equilibria depend, again, on $\delta$ and $\kappa_p$. When it is unlikely to discover a vaccine or cure (low $\delta$), the early response to the pandemic critically depends on whether COVID-19 becomes endemic. If it becomes endemic (benchmark and $\kappa_q=0.25\kappa_p$), the social planner restricts social activity further as the present value of the costs of the endemic steady-state are larger. But if COVID-19 does not become endemic ($\beta_q=0.25\beta_p$ or  $\sigma=0.25$), the social planner is more lenient. A low cost of infection of primary agents, $\kappa_p$, grants room for maneuver for the social planner to act early to endemic steady-states due to the curvature of the utility function. Therefore,  mandated social-distancing visibly increases with the overall costs of the pandemic in Panel C of Figure \ref{fig_robust2}. \\

The optimal behavior of secondary susceptible agents is much different from that of primary susceptible agents irrespective of $\delta$ and $\kappa_p$. If it is unlikely that secondary agents are reinfected ($\beta_q$ is low), they restrain social activity by much less than primary ones, which is problematic from a social perspective because they expose other agents (especially primary) significantly. Thus, even if susceptible agents are unlikely to be reinfected, policymakers should be aware that these agents are likely to be excessively active.\\

This problem of excessive social activity in the decentralized equilibrium is even worse if $\kappa_q=0.25\kappa_p$. As agents are not altruistic, they only care about their own risks. A lower cost of reinfection then significantly lowers their incentives to social-distance. In contrast, the social planner would like secondary agents to substantially constrain their activity because their viral shedding and probability of infection are unchanged and many susceptible agents are still primary susceptible.\footnote{In this regard, secondary agents are similar to young agents in models that breakdown agents based on age \citep{Acemoglu_Chernozhukov_Werning_Whinston_covid2020, Gollier_covid2020}. In those models, because young agents know that they are less likely to suffer if infected, they are too active from a social perspective as they increase exposure of older individuals.} The scenario of $\kappa_q=0.25\kappa_p$ also shows that agents asymptotically constrain social activity, even in the decentralized equilibrium, because COVID-19 becomes endemic and the costs of infection remain high (these costs imply a probability of dying of 0.16\% in the benchmark). If these costs were lower, closer to those of endemic human coronaviruses, agents in a decentralized equilibrium would behave almost as if there was no virus which is what we observed until the COVID-19 pandemic.\\

The results are very different if $\sigma=0.25$. Recall that $\sigma$  measures how likely secondary infected shed virus onto susceptible. Since $\sigma$ pertains only to the externality caused by secondary agents' actions,  it does not affect decisions in the decentralized equilibrium: secondary susceptible agents act  as primary susceptible agents. A social planner, in contrast, would allow secondary agents to enjoy relatively  more social activity. Both primary and secondary agents, however, benefit indirectly from the lower viral-shedding of secondary infected agents, which allows them to enjoy more social activity, converging asymptotically to full social activity in both equilibria.

\section{What If It Was Today?}\label{sec_today}
Following the SARS-CoV2 outbreak, governments around the world have combined several NPIs to change the natural course of the pandemic. To account for this change,  in this section, we base our simulations on initial conditions matching  the current (epidemiological) state of the COVID-19 pandemic. \\

In the (new) initial conditions, we accommodate a compromise between the epidemiological state in the US and four European countries,  France, Italy, Spain, and the UK, as of 1 July 2020. On this day, the fraction of (currently) infected population  was approximately 0.46\% in the US;  0.09\% in France and 0.02\% in Italy.\footnote{Statistics consulted in \href{https://coronavirus.jhu.edu/map.html}{https://coronavirus.jhu.edu/map.html} on 2 July 2020.} These numbers are likely to be understated as authorities fail to test and identify many of infected and especially asymptomatic people (see references in \citealp{Stock_covid2020} for evidence on the proportion of asymptomatic). Bearing in mind the understatement and cross-country differences in the numbers,  we find a compromise at $i=0.2\%$. To set the initial number of recovered agents, we look at the evidence from antibody surveys. In France, Spain, and the UK,  antibody surveys suggest that slightly more than 5\% of the population has antibodies against SARS-CoV-2.\footnote{See the ONS COVID-19 Infection Survey for the UK; for France and Spain, see \cite{Salje_covid2020} and \cite{Pollan_covid2020}.} Given that the fraction of infected population ratio is two to three times higher in the US than in France,  Spain, and the UK, we find a compromise at $r=6\%$.\\

In all countries that we examined for this section, identified infected individuals are quarantined. This NPI naturally reduces contagion and we model it as an exogenous reduction in the social activity of some infected agents. In particular, we assume that 50\% of infected agents, which is within the current estimated range of asymptomatic cases, are identified and cannot enjoy maximum social activity. In case infected individuals are identified, they enjoy 40\% of normal social activity, which increases the expected costs of infection. Thus, average social activity of infected individuals falls by 30\%. Other NPIs, like mandatory mask use, differ across countries. In France and the UK,  mask use is only mandatory in public transport, whereas in Spain, it is mandatory even in open-air spaces if it is not possible to maintain physical distance.\footnote{At the time that we write this paper, France and the UK have announced mandatory mask use in shops.}  In our model, we treat mask use (mandatory or not) as an exogenous reduction in contagiousness, $\beta_p$ and $\beta_q$, by 30\%. In sum, these NPIs reduce $\beta_p$ and $\beta_q$ by slightly over 50\%.\footnote{Crucially, $R0$ is still above one as the pandemic would asymptotically disappear if $R0<1$. But $R0$ permanently below one seems unlikely as pointed by the second wave of infections in Australia and South Korea.}\\

\begin{figure}[!htbp]
	\centering
	\caption{\linespread{1.0}\selectfont 
		\label{fig_today}%
		What if It Was Today?
		\vspace*{0.1cm}
	}
	\includegraphics[scale=0.7]{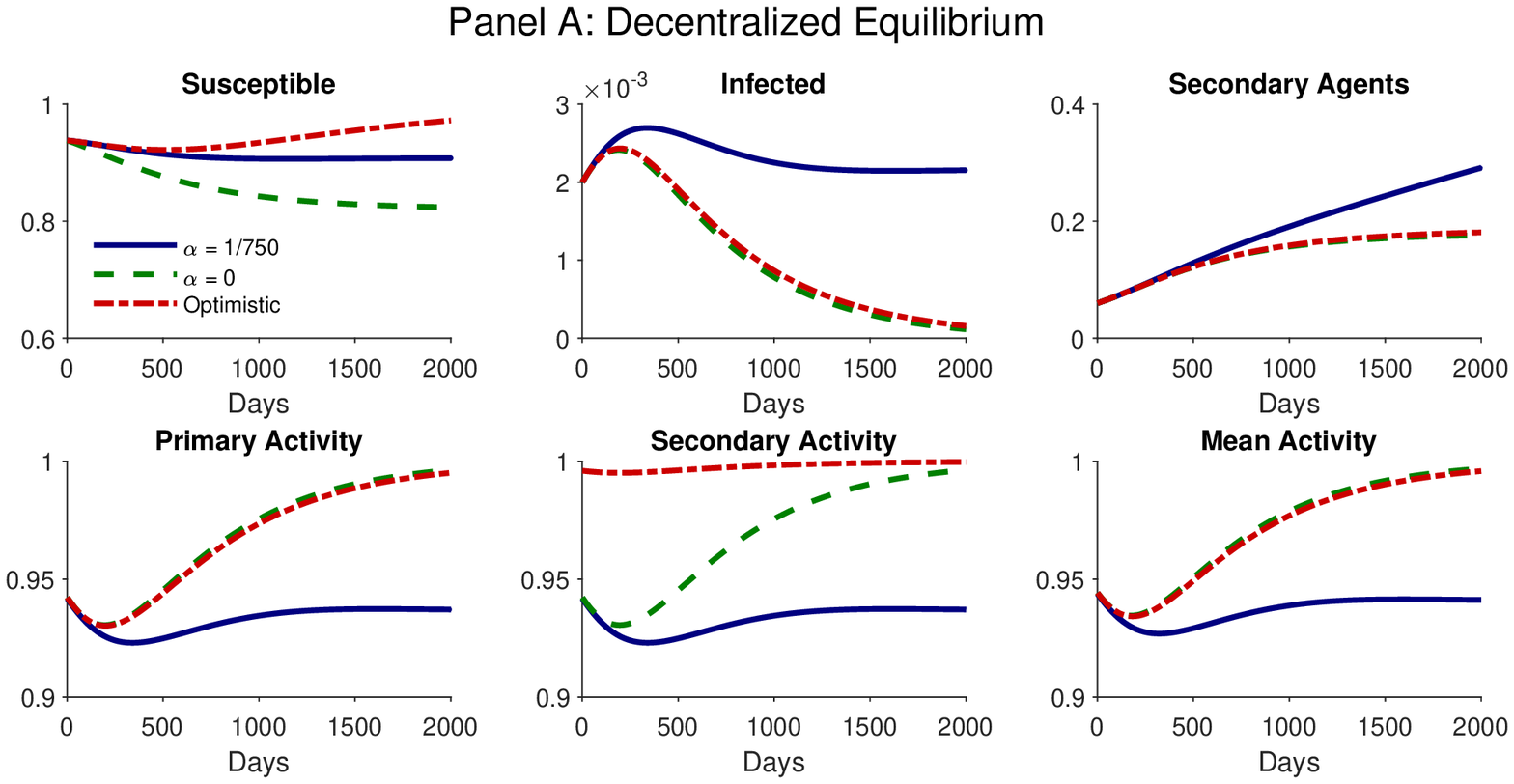}\vspace{0.35cm}
	\includegraphics[scale=0.7]{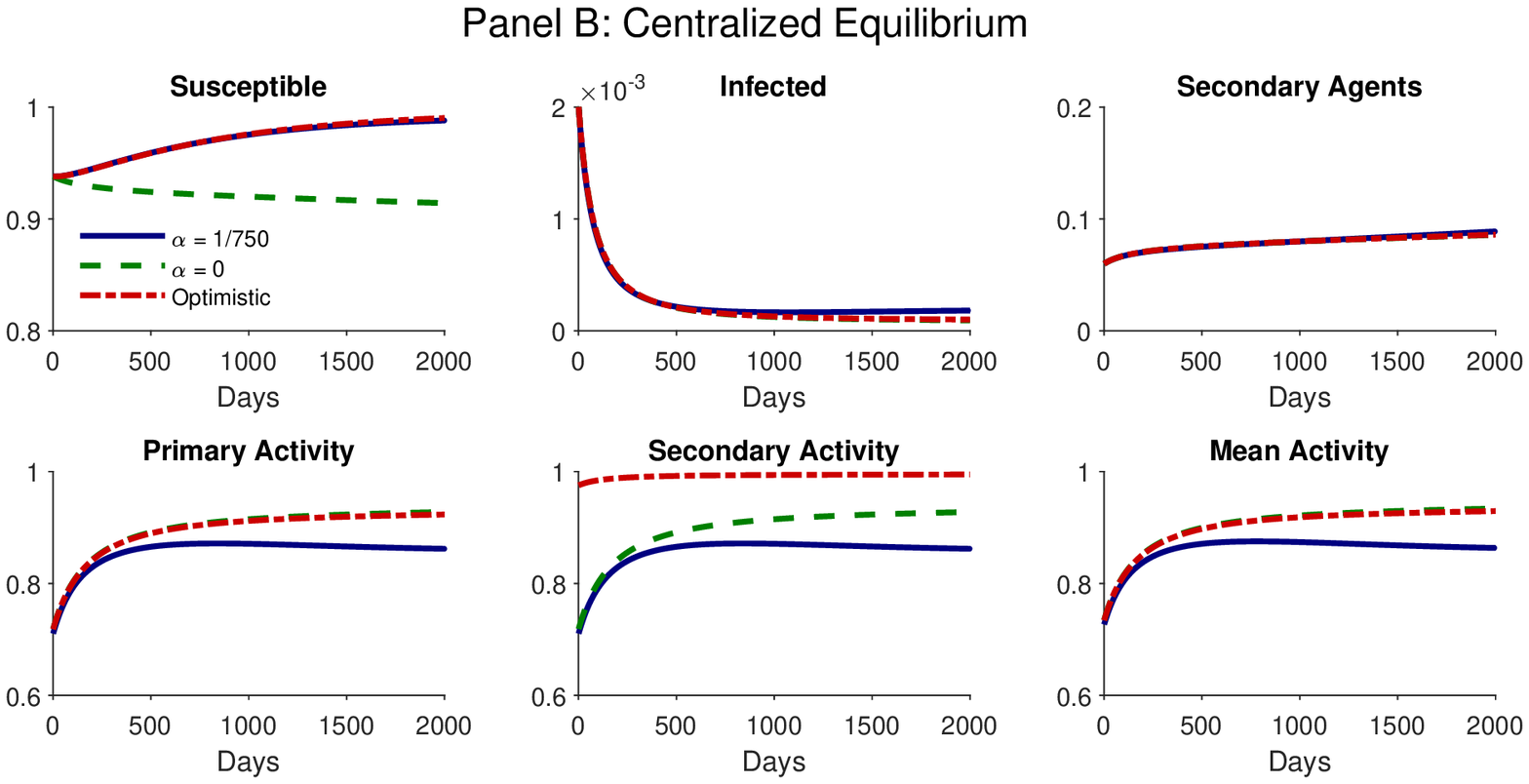}
	\parbox{13cm}%
	{\vspace*{.5cm}
		\footnotesize \textit{Note:} Optimistic refers to the case in which $\beta_q=0.25\beta_p$, $\sigma=0.25$, and $\kappa_q=0.25\kappa_p$. Susceptible agents are $s_{p,t} + s_{q,t}$; infected agents are $i_{p,t} + i_{q,t}$; secondary agents are $s_{q,t} + i_{q,t} + r_t$; primary activity is $a_{p,t}$; secondary activity is $a_{q,t}$; and mean activity is $s_{p,t}a_{p,t} + s_{q,t}a_{q,t} + 0.7(i_{p,t} + i_{q,t}) + r_t$.
	}%
\end{figure}

\begin{figure}[!htbp]
	\centering
	\caption{\linespread{1.0}\selectfont 
				\label{fig_today2}%
		What if It Was Today? - Centralized Equilibria
		\vspace*{0.1cm}
	}
	\includegraphics[scale=0.7]{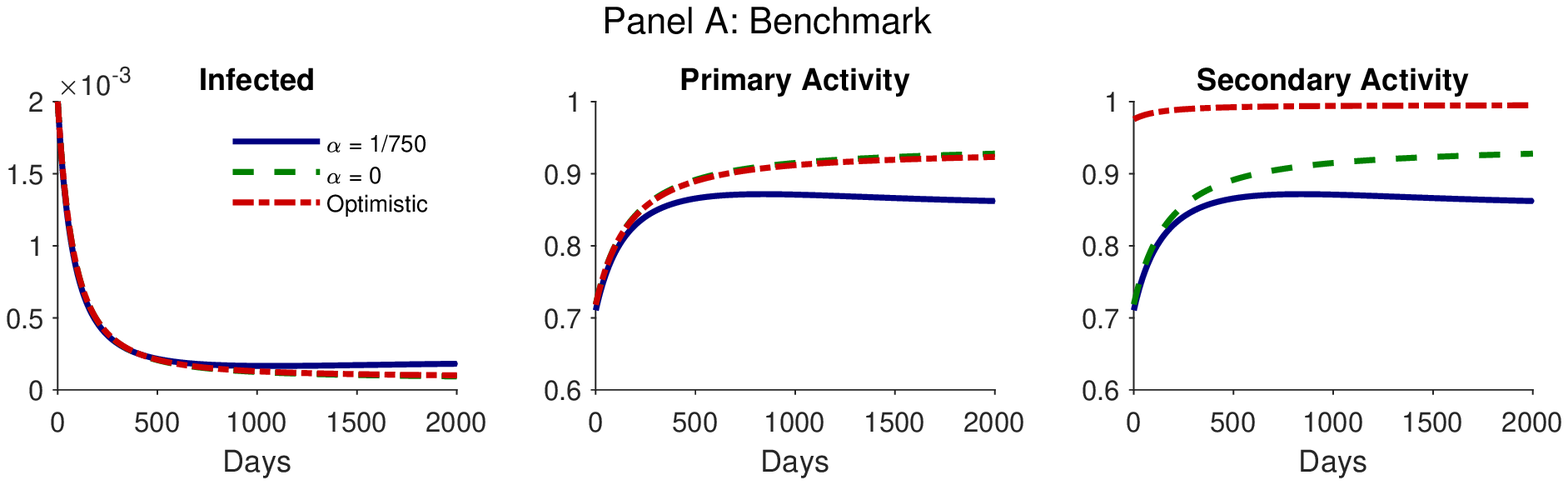}\vspace{0.35cm}
	\includegraphics[scale=0.7]{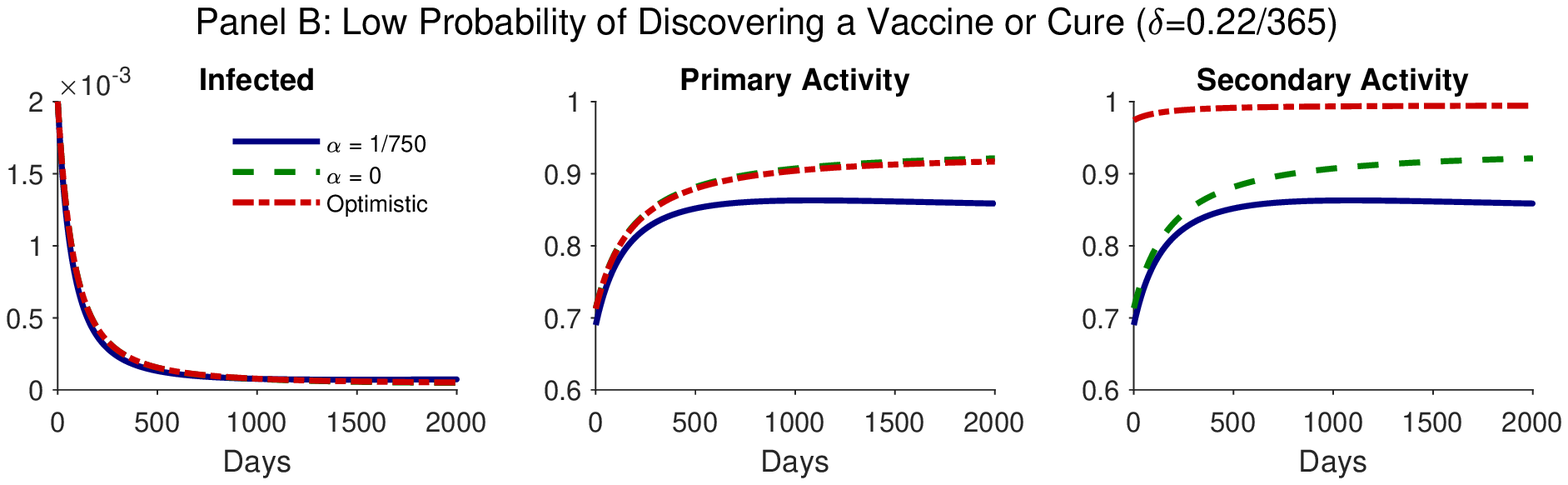}\vspace{0.35cm}
	\includegraphics[scale=0.7]{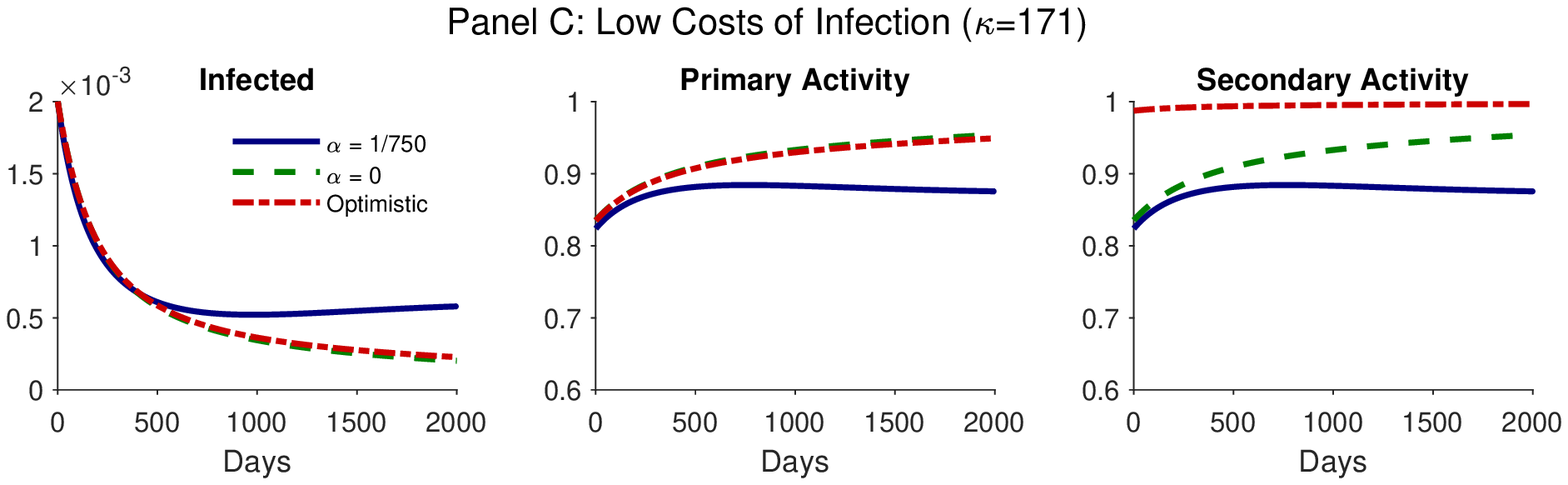}
	\parbox{13cm}%
	{\vspace*{.5cm}
		\linespread{1.0}\selectfont 
		\footnotesize \textit{Note:} Optimistic refers to the case in which $\beta_q=0.25\beta_p$, $\sigma=0.25$, and $\kappa_q=0.25\kappa_p$. Infected agents are $i_{p,t} + i_{q,t}$; primary activity is $a_{p,t}$; secondary activity is $a_{q,t}$.
	}%
\end{figure}

We depict the results in Figures \ref{fig_today} and \ref{fig_today2}. Blue (solid) lines assume the benchmark values for the the rest of the parameters. Green (dashed) lines assume that agents are permanently immune. Red (dot-dashed) lines assume that in their  contagiousness and cost of infection, secondary agents differ substantially from primary agents: $\beta_q=0.25\beta_p$, $\sigma=0.25$, and $\kappa_q=0.25\kappa_p$.\footnote{In these simulations, we assume that the initial fraction of secondary susceptible and infected individuals is zero.} Compared to our previous simulations, the other NPIs significantly elevate social activity because of the fall in contagiousness.  Furthermore, the simulations suggest that individuals and policymakers do not need to know the duration of immunity and how secondary agents differ from primary ones until at least  2021 even if $\delta$ and $\kappa_p$ are low.\footnote{Although there are slightly visible differences in terms of optimal primary activity if $\delta$ and $\kappa_p$ are low, the implied dynamics of infected individuals is almost unchanged. Optimal secondary activity depends much more on the scenario for waning immunity, but there are very few agents that are secondary susceptible.}  Thus, the combination of lower contagiousness and relatively high initial infections reduce the relevance of waning immunity  even in centralized equilibria, making the the social planner less responsive to future infection costs. This suggests that if current NPIs remain in place, there is still substantial time to learn about the duration of immunity. Yet, given the implications of the mortality rate for social activity, and consistent with \cite{Hall_Jones_Klenow_covid2020}, learning about the actual infection-fatality rate seems highly important.

\section{Concluding Remarks}\label{sec_conclusion}
It is likely that immunity against COVID-19 eventually wanes and those that are immune today will face the risk of a reinfection \citep{Edridge_covid2020, Huang_covid2020, Kellam_Barclay_covid2020, Seow_etal_covid2020}. This scenario is especially problematic if COVID-19 becomes endemic as other human endemic coronaviruses. We show that if COVID-19 reaches an endemic steady-state and a vaccine or cure is not discovered, social-distancing is here to stay. But, on the bright side, we also show that optimal decentralized and centralized choices do not necessarily depend on waning immunological memory for many months following the initial outbreak/contagion. This is especially the case if a vaccine is expected soon, the costs of infection are already large in the short run, and other NPIs that lower contagiousness remain in place. Before making irreversible decisions, individuals and policymakers seem to have time to learn more about immunological memory against SARS-CoV-2 and answer the call for serological studies from \cite{Kellam_Barclay_covid2020}, \cite{Kissler_etal_covid2020}, and \cite{Lerner_etal_covid2020}.\\

Yet, in 6-12 months, we do need to know more about how antibodies and T-cells defend the human body against SARS-CoV-2. In particular, we must know how long immunity lasts and whether individuals that were infected (secondary agents) differ substantially from those that were never infected (primary agents). The longer immunity lasts, the less demanding should social-distancing be. And, in the limit, if immunity lasts a lifetime, then COVID-19 does not reach an endemic steady-state and social-distancing will sooner or later be unwarranted. Furthermore, if secondary agents may be reinfected but are somewhat protected against the virus, then COVID-19 may not become endemic. Yet, the way in which secondary agents differ from primary agents is crucial to design policy. For example, if most of the gains from the additional protection are private -- because secondary agents are less likely to die or less likely to be reinfected -- then secondary agents are excessively active from a social viewpoint. If, on the other hand, most of the gains from the additional protection are social -- because secondary agents shed less virus -- then the decentralized and centralized equilibria are closer and less social-planning is required.\\

Even though most of the economics literature assumes permanent immunity, this simplification may not have dire consequences in the short run. If a vaccine or cure is expected soon, the costs of infection are not small, and other NPIs are in place, then our model suggests that the optimal response in the initial months of the pandemic is virtually independent of waning immunity. The same is true if secondary agents, despite no longer immune, develop a strong protection against SARS-CoV-2 or shed much less virus. But, if these conditions do not hold, many of the policy prescriptions need to be revised as they rely on the possibility of herd immunity.

\newpage
\appendix
\section{Robustness Checks of Decentralized Equilibria}
\setcounter{figure}{0}
\renewcommand{\thefigure}{A\arabic{figure}}
\begin{figure}[!htbp]
	\centering
	\caption{\linespread{1.0}\selectfont 
			\label{fig_7}%
		The Role of Immunity Duration - Decentralized Equilibria
		\vspace*{0.1cm}
	}
	\includegraphics[scale=0.7]{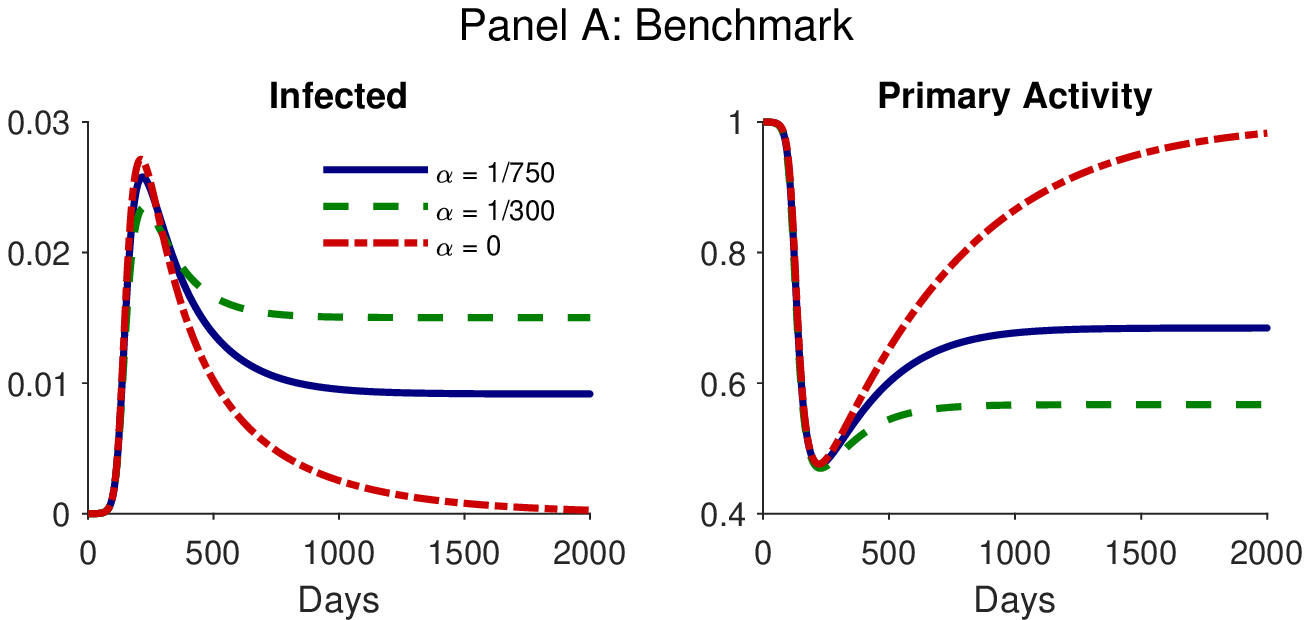}\vspace{0.35cm}
	\includegraphics[scale=0.7]{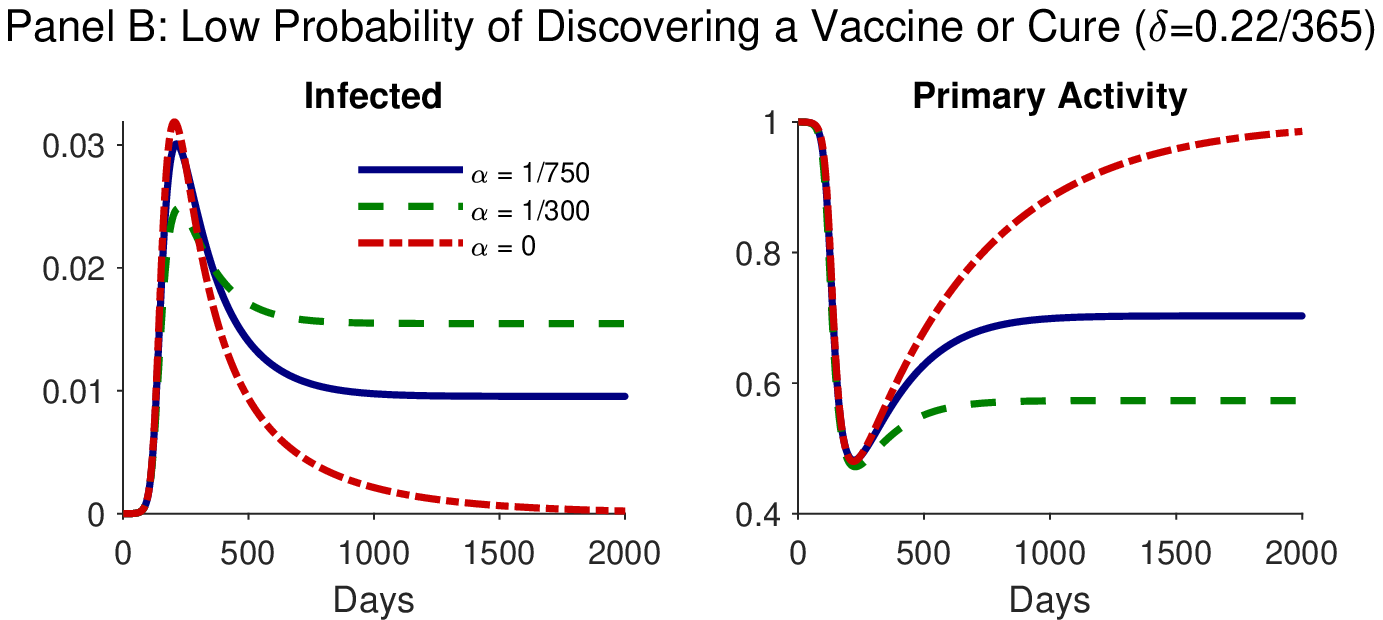}\vspace{0.35cm}
	\includegraphics[scale=0.7]{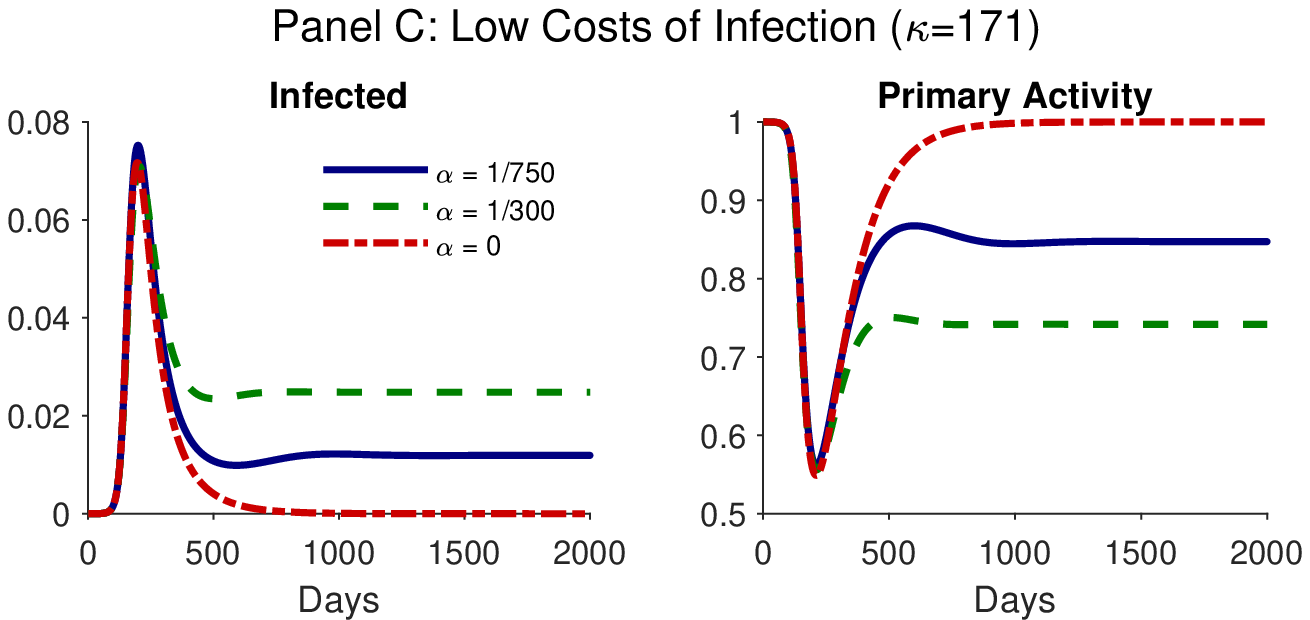}
	\parbox{13cm}%
	{\vspace*{.5cm}
		\linespread{1.0}\selectfont 
		\footnotesize \textit{Note:} Infected agents are $i_{p,t} + i_{q,t}$; primary activity is $a_{p,t}$ (which, in this case, equals secondary activity, $a_{q,t}$).
	}%
\end{figure}

\begin{figure}[!htbp]
	\centering
	\caption{\linespread{1.0}\selectfont 
		What if Primary and Secondary Agents Differ? - Decentralized Equilibria
		\vspace*{0.1cm}
	}
	\includegraphics[scale=0.7]{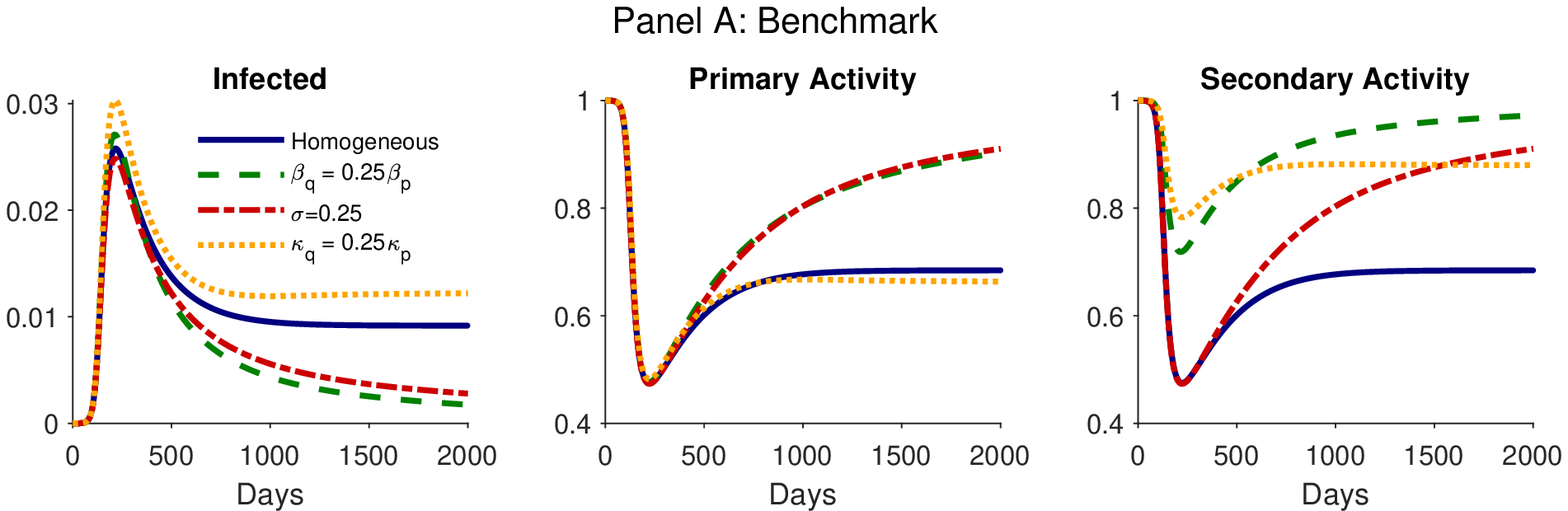}\vspace{0.35cm}
	\includegraphics[scale=0.7]{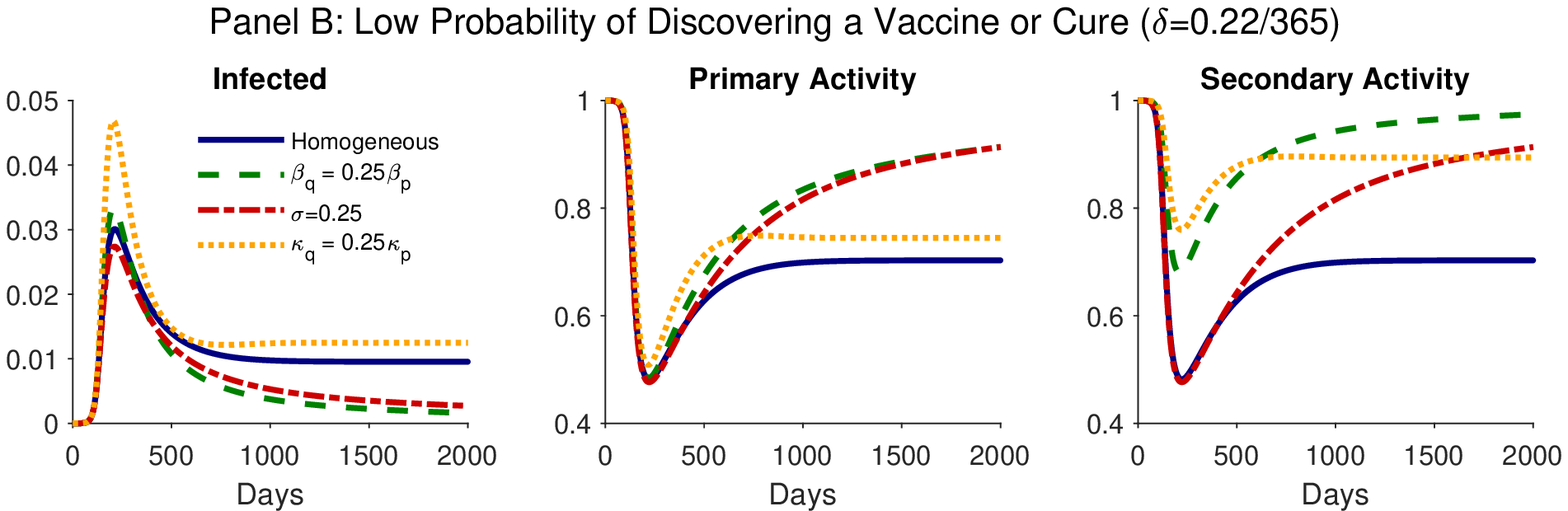}\vspace{0.35cm}
	\includegraphics[scale=0.7]{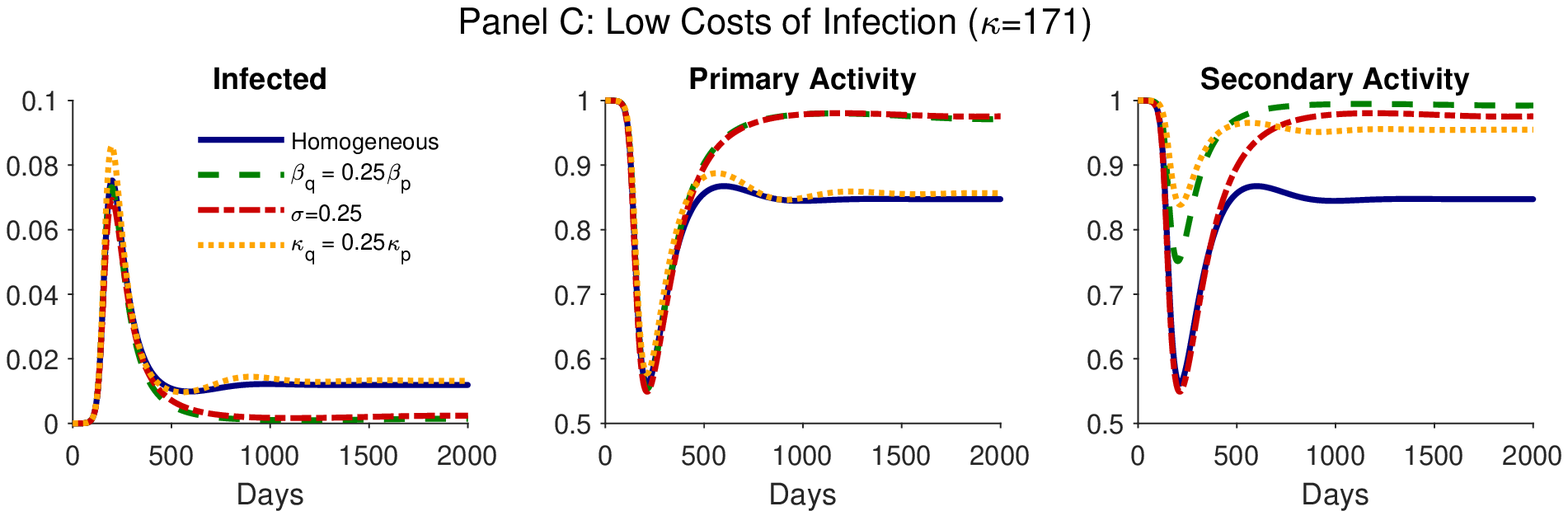}
	\parbox{13cm}%
	{\vspace*{.5cm}
		\linespread{1.0}\selectfont 
		\footnotesize \textit{Note:} Homogeneous refers to the case in which secondary and primary agents are alike. Infected agents are $i_{p,t} + i_{q,t}$; primary activity is $a_{p,t}$; secondary activity is $a_{q,t}$.
	}%
\end{figure}

\begin{figure}[!htbp]
	\centering
	\caption{\linespread{1.0}\selectfont 
		What if It Was Today? - Decentralized Equilibria
		\vspace*{0.1cm}
	}
	\includegraphics[scale=0.7]{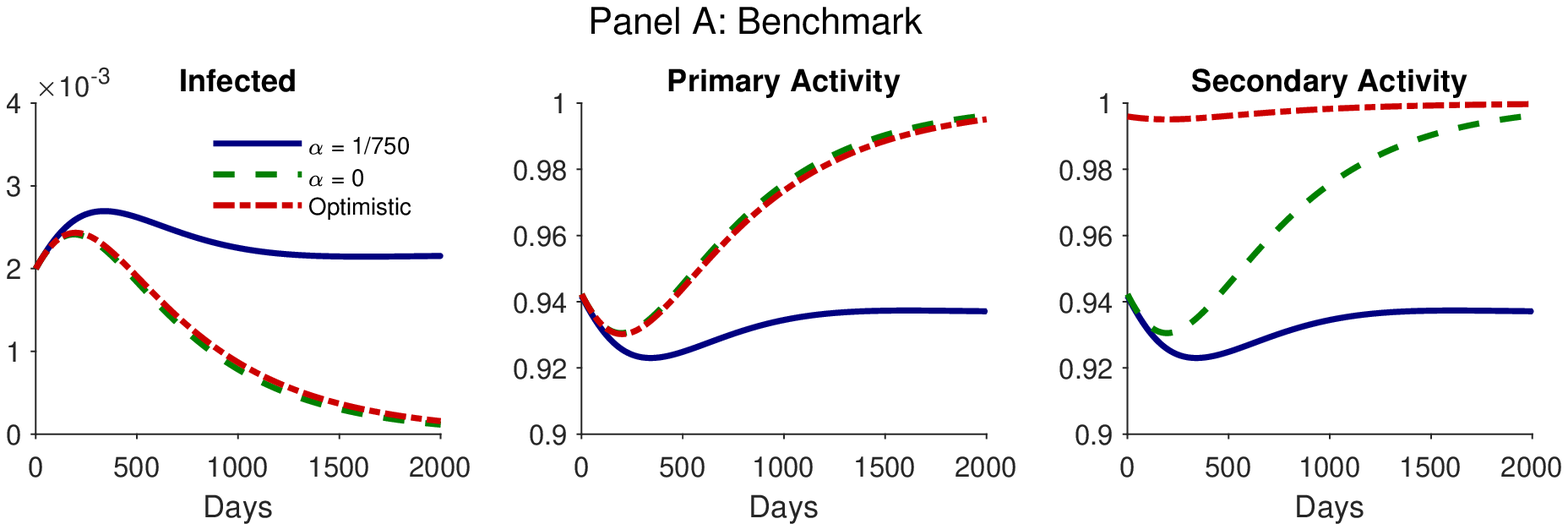}\vspace{0.35cm}
	\includegraphics[scale=0.7]{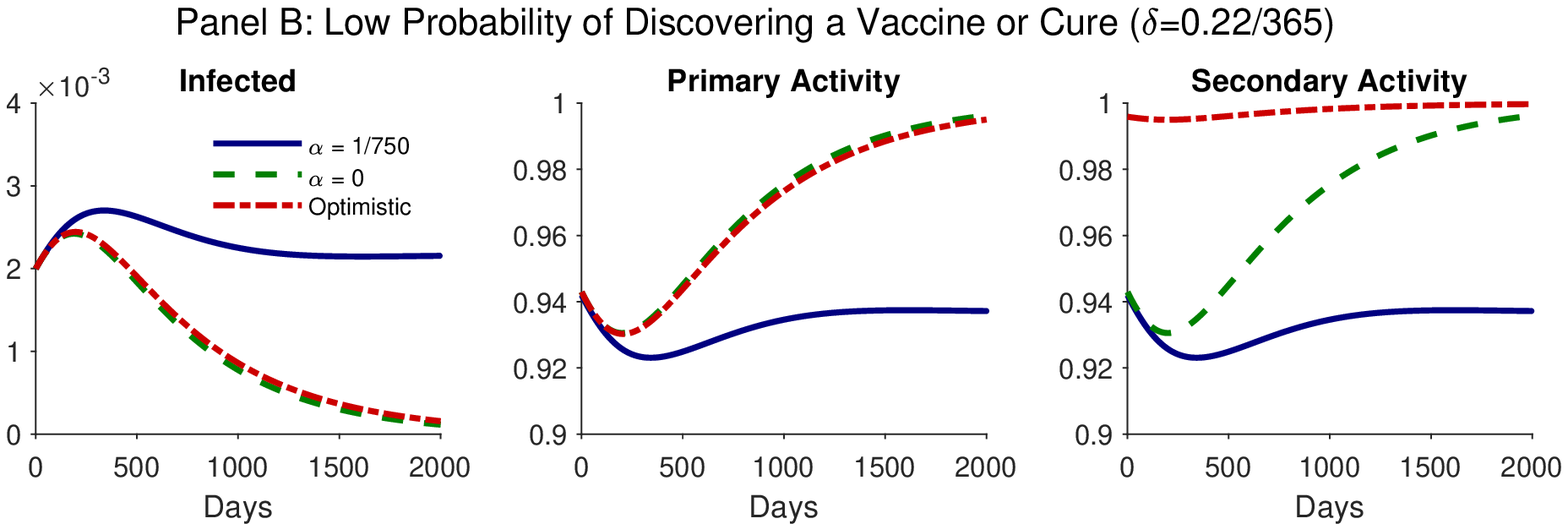}\vspace{0.35cm}
	\includegraphics[scale=0.7]{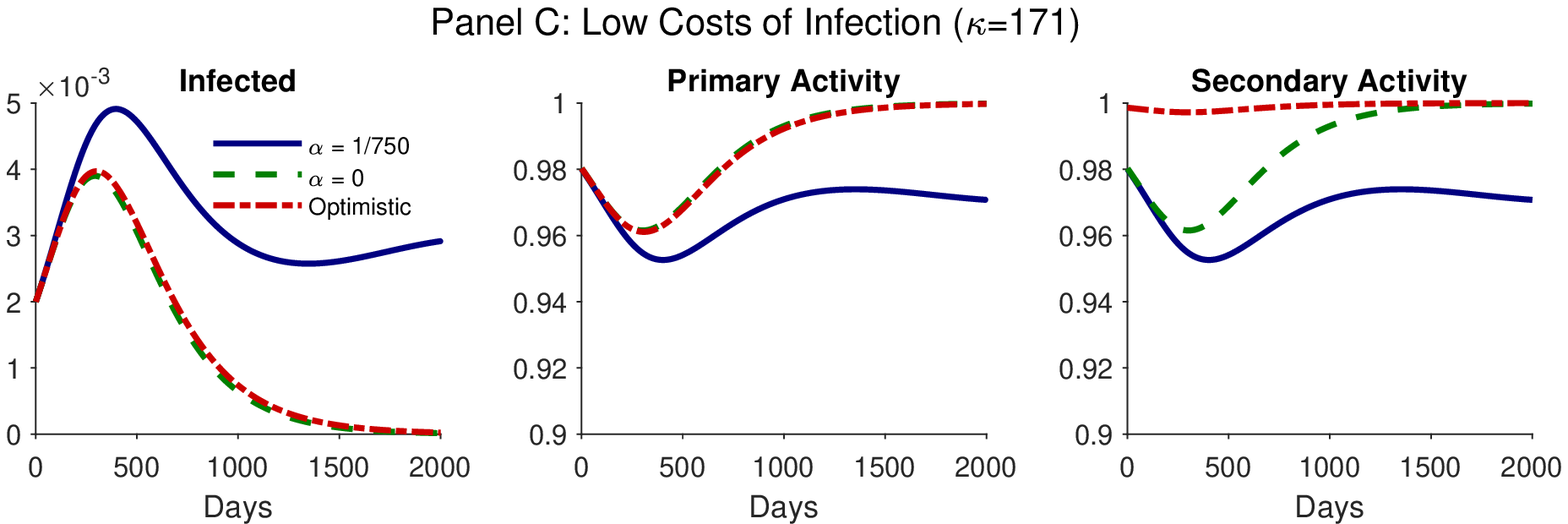}
	\parbox{13cm}%
	{\vspace*{.5cm}
		\linespread{1.0}\selectfont 
		\footnotesize \textit{Note:} Optimistic refers to the case in which $\beta_q=0.25\beta_p$, $\sigma=0.25$, and $\kappa_q=0.25\kappa_p$. Infected agents are $i_{p,t} + i_{q,t}$; primary activity is $a_{p,t}$; secondary activity is $a_{q,t}$.
	}%
\end{figure}

\end{spacing}
\end{document}